\documentclass[aps,prl,groupedaddress,superscriptaddress, twocolumn
,amsmath,amssymb]{revtex4-1}
\usepackage{romannum}
\usepackage{graphicx}
\usepackage{bm}
\usepackage{ulem}

\begin{document}

\title{3D Quantum Hall Effect Manipulated by Chiral Landau Levels in Weyl Semimetals}
\author{Hailong Li}
\affiliation{International Center for Quantum Materials, School of Physics,
Peking University, Beijing 100871}
\author{Haiwen Liu}
\affiliation{Center for Advanced Quantum Studies, Department of Physics,
Beijing Normal University, Beijing 100875}
\author{Hua Jiang}
\email{jianghuaphy@suda.edu.cn}
\affiliation{School of Physical Science and Technology, Soochow University, Suzhou 215006, China}
\affiliation{Institute for Advanced Study, Soochow University, Suzhou 215006, China}
\author{X. C. Xie}
\email{xcxie@pku.edu.cn}
\affiliation{International Center for Quantum Materials, School of Physics,
Peking University, Beijing 100871}
\affiliation{Beijing Academy of Quantum
Information Sciences, Beijing 100193, China}
\affiliation{CAS Center for
Excellence in Topological Quantum Computation, University of Chinese Academy
of Sciences, Beijing 100190, China}
\date{\today }

\begin{abstract}
We investigate the 3D quantum Hall effect in Weyl semimetals and elucidate a global picture of the edge states. The edge states hosting 3D quantum Hall effect are combinations of Fermi arcs and chiral bulk Landau levels parallel to the magnetic field. The Hall conductance, $\sigma_{xz}^H$, shows quantized plateaus at Weyl nodes while tuning the magnetic field. However, the chiral Landau levels manipulate the quantization of Weyl orbits, especially under a tilted magnetic field, and the resulting edge states lead to distinctive Hall transport phenomena. A tilted magnetic field contributes an intrinsic initial value to $\sigma_{xz}^H$ and such initial value is determined by the tilting angle $\theta$. Particularly, even if the perpendicular magnetic field is fixed, $\sigma_{xz}^H$ will change its sign with an abrupt spatial shift of edge states when $\theta$ exceeds a critical angle $\theta_c$ in an experiment. Our work uncovers the unique edge-state nature of 3D quantum Hall effect in Weyl semimetals.
  \end{abstract}

  \maketitle

  \textit{Introduction.---}
  Weyl semimetals are 3D topological quantum materials of which bulk energybands are gapped except for even number of discrete points in the momentum space, named Weyl nodes~\cite{reviewRMP,wanxiangangPRB,xugangPRL,burkovPRL,balentsPRB,daixiPRX,hasanNC,dinghongPRX,liuenkeNP}. The bulk energy dispersion near a Weyl node is linear and can be described by the Weyl equation. At the surfaces of a Weyl semimetal, there exist topologically protected surface states, so-called Fermi arcs, and they connect Weyl nodes with opposite chiralities.
  The nontrivial band structure has been observed~\cite{diracSCIENCE2014,xusuyangSCIENCE02,diracPRL2014,dinghongPRX,xusuyangSCIENCE}.
  Due to the topological electronic structure, Weyl semimetals can induce an exotic phenomenon known as the chiral anomaly~\cite{chiralanomalyPLB,chiralanomalyPRB01,chiralanomalyPRB02}. Here, a strong magnetic field drives bulk states into the chiral Landau levels, where the velocities at Weyl nodes of different chiralities are opposite. Further, 3D quantum Hall effect (QHE) is theoretically predicted to occur in Weyl semimetals, where the Fermi arcs at the top and bottom surfaces form a closed loop via a ``wormhole'' tunneling assisted by the Weyl nodes~\cite{luhaihzouPRL}. Meanwhile, exotic quantum Hall phenomena under a magnetic field are observed in topological semimetals ~\cite{nagaosaNC,xiufaxianNC,liaozhiminPRL,xiufaxianNATURE}. However, a global picture that how the edge states are evolved and form a closed trajectory is still missing in 3D QHE. Particularly, edge states should emerge on the side surfaces. Some side surfaces are topologically nontrivial in Weyl semimetals, while others are trivial~\cite{reviewRMP}. Thus, such edge states are not describable by the conventional theory for the Weyl orbits.

  \begin{figure}[t!]
    \includegraphics[width=\columnwidth]{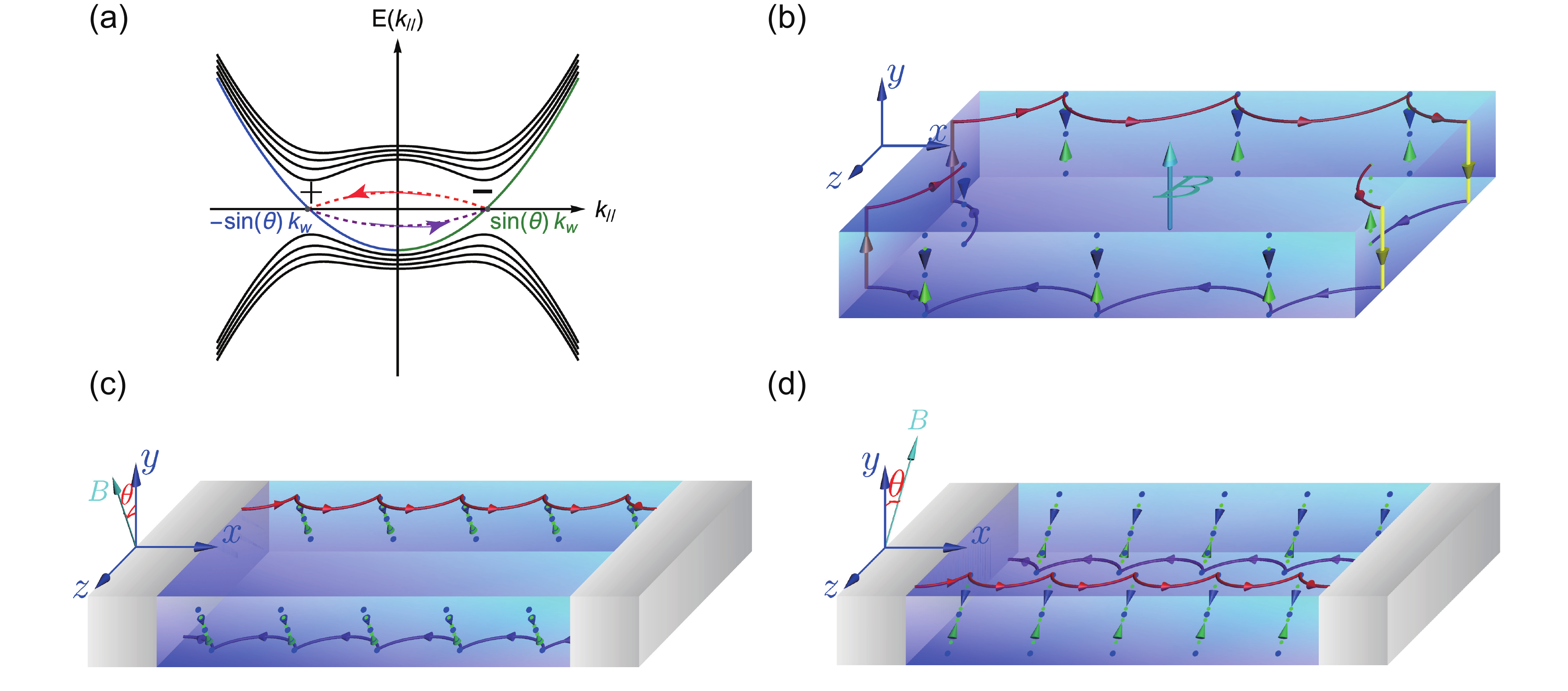}
    \caption{(a) The schematic plot of the energyband structure under a tilted magnetic field $\bm{B}$. The two weyl nodes are distributed at $(0, 0, \pm k_w)$ without $\bm{B}$. $\theta$ stands for the angle of $\bm{B}$ from the surface normal and $k_\parallel$ is the wavevector parallel to $\bm{B}$. The red and purple dashed lines represent the Fermi-arcs on the top and bottom surfaces, respectively. The blue and green solid lines represent the chiral bulk landau levels.[(b)--(d)] Physical pictures of the edge states with $\theta=0$, $\theta>0$ and $\theta<0$. The red and purple curves stand for the trajectories on the top and bottom surfaces, respectively. The dashed blue and green lines are the bulk chiral Landau level states with opposite velocities parallel to $\bm{B}$.  The brown and yellow lines are from Fermi arc states at side surfaces. The top, front and left surfaces are located at $y=L_y/2$, $z=L_z/2$ and $x=-L_x/2$, respectively.
    }
    \label{Fig1}
  \end{figure}

  In this Letter, our aim is to figure out the edge states trajectories and how they are manipulated by a tilted magnetic field $B$ [Fig.~\ref{Fig1}]. Under a perpendicular magnetic field, a closed 3D trajectory of edge states is theoretically proposed [Fig.~\ref{Fig1}(b)]. In the $x$-$y$ plane, the chiral Landau levels assist the semi-circle of the Fermi-arc states on the top(bottom) surface to form the skipping edge states. In the $y$-$z$ plane, both the chiral Landau levels and the Fermi-arc states on the side surfaces are involved to form the closed trajectory. Furthermore, we find the quantum channels of edge states can be manipulated by a tilted magnetic field $\bm{B}$, since involved chiral Landau levels are parallel to $\bm{B}$ [Fig.~\ref{Fig1}(c)-(d)]. Consequently, the distribution of the edge states and behavior of the Hall conductance are significantly changed. When tuning $B_z$, even for a constant $B_y$, the edge states can change their location from one side to the other, and subsequently the initial value of Hall conductance changes its sign by rotating $\bm{B}$. These distinctive phenomena cannot be attributed to the gauge potential of the magnetic field, but determined by the intrinsic topological nature of Weyl semimetals, which are closely related to several experimental candidates.

  \textit{Model and Methods.---}
We adopt a $2\times2$ two-node minimal model to describe a 3D Weyl semimetal~\cite{luhaihzouPRL,luhaizhouPRB,shenshunqingBOOK,murakamiPRB},
  \begin{equation}
    \begin{aligned}
      H(\bm{k})=& D_{1} k_{y}^{2}+D_{2}\left(k_{x}^{2}+k_{z}^{2}\right)+A\left(k_{x} \sigma_{x}+k_{y} \sigma_{y}\right) \\ &+M\left(k_{w}^{2}-\bm{k}^{2}\right) \sigma_{z} \end{aligned}
      \label{eq1}
    \end{equation}
where $D_1, D_2, A, M, k_w$ are parameters and $(\sigma_x, \sigma_y, \sigma_z)$ are Pauli matrices. The energy dispersion of this model is $E_{\pm}(\bm{k})=D_{1} k_{y}^{2}+D_{2}\left(k_{x}^{2}+k_{z}^{2}\right)\pm\sqrt{M^2(k_w^2-\bm{k}^2)^2+A^2(k_x^2+k_z^2)}$. Therefore, ``$-$'' and ``$+$'' Weyl nodes are separately located at $(0,0,k_w)$ and $(0,0,-k_w)$ with energy $E_0=D_2k_w^2$. Hereafter, the Fermi energy is fixed at Weyl nodes, i.e. $E_F=E_0$. Besides, $D_1$ and $D_2$ terms brings curved Fermi arcs and a finite area of Fermi surface, $S_{\bm{k}}$, in the Brillouin zone \cite{luhaihzouPRL}. Nonzero $S_{\bm{k}}$ makes the Fermi-arc contribution reflected in the Hall conductance as the magnetic field changes and is more suitable for real materials. In the presence of an external magnetic field $B$, the Peierls substitution is required\cite{Peierls}. Then, we discretize the Hamiltonian into a cubic lattice model, and use the nonequilibrium Green's function (NEGF) method to calculate the local density of states (LDOS), the local current density and the Hall conductance~\cite{SM,negfPRB1994,jianghuaPRB,numerical01,numerical01,numericalo3}.

To describe trajectories of the electrons in real space, we utilize the semicassical equations of motion~\cite{niuqianRMP,acpotterNC}:
\begin{equation}
  \begin{aligned}
    \dot{\boldsymbol{r}} &=\frac{1}{\hbar} \frac{\partial \boldsymbol{\varepsilon}_{n}(\boldsymbol{k})}{\partial \boldsymbol{k}}
   \\
    \hbar \dot{\boldsymbol{k}} &=
    -e \dot{\boldsymbol{r}} \times \boldsymbol{B}(\boldsymbol{r}).
  \end{aligned}\label{eq2}
\end{equation}
Here, we ignore the Berry curvature term~\cite{berrycurvature_expla}. Eq.~\ref{eq2} establishes a map between the momentum ${\boldsymbol{k}}$ and the position ${\boldsymbol{r}}$~(Sec. S1 of \cite{SM}).
Electrons on the topological surfaces are from the Fermi arcs [red and purple curves in Fig.~\ref{Fig1}]. Therefore, $\dot{\bm{r}}$ of the electrons is orientated along the normal of Fermi arcs and $\dot{\bm{k}}$ is tangent to the arcs. In $\boldsymbol{k}$ space, electrons slide along Fermi arcs and map a trajectory into the real space. As for the electrons in the bulk or trivial surfaces, they are from the chiral Landau levels [blue and green lines in Fig.~\ref{Fig1}]. Thus, it makes $\dot{\bm{k}}$ zero that $\dot{\bm{r}}$ is parallel to $\bm{B}$. Trajectories in real space are straight lines parallel to $\bm{B}$. Next, we discuss the semiclassical picture of edge states in detail.

  \begin{figure}[b!]
    \centering
    \includegraphics[width=\columnwidth]{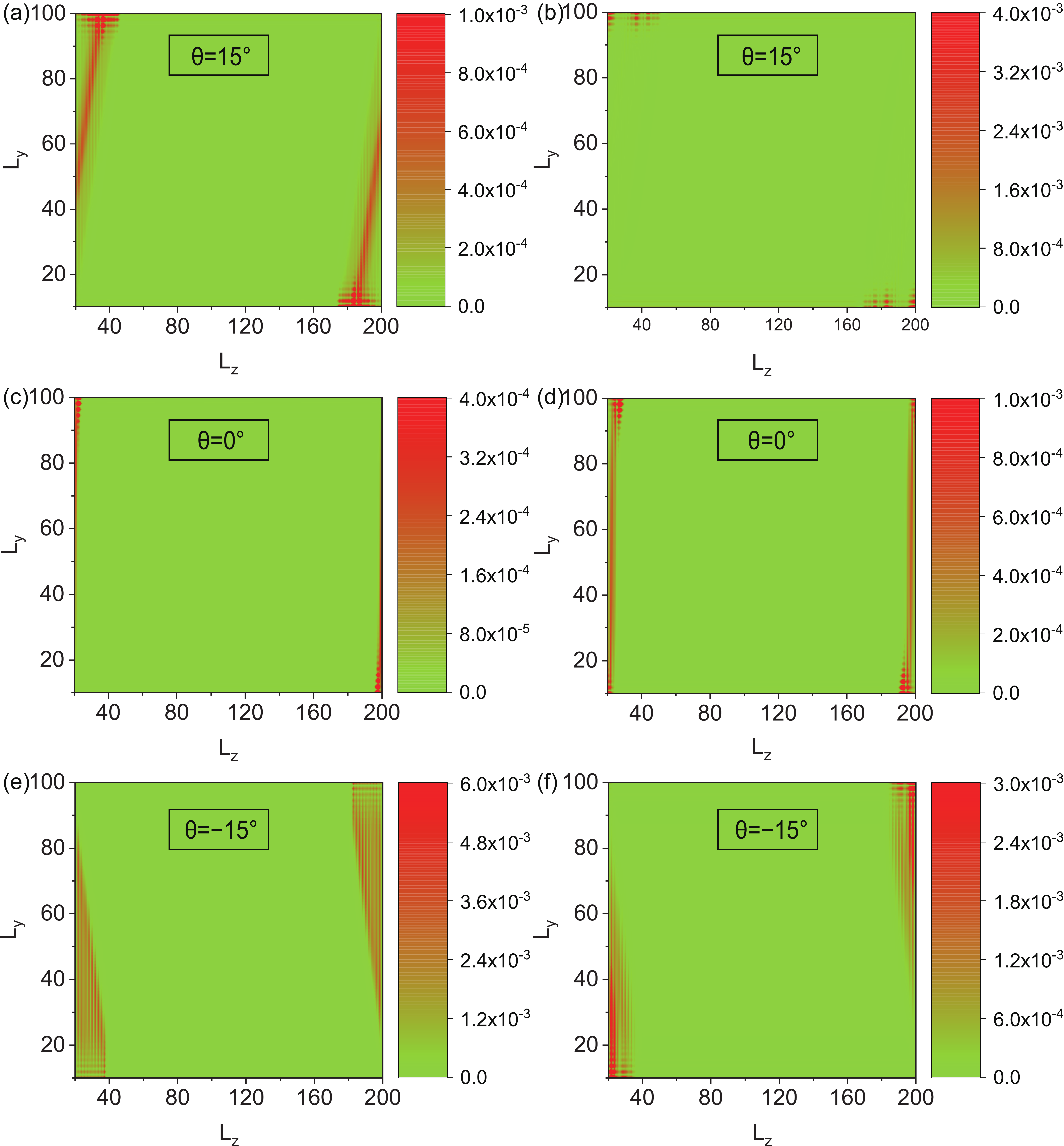}
    \caption{LDOS on the y-z cross section for an infinite Weyl semimetal along the $x$ direction. The Fermi energy is fixed at Weyl nodes. (a)-(f) LDOS under fixed $B_y=1/20$ (a),(c),(e) and $B_y=1/44$ (b),(d),(f). 
    The size of the cross section is $L_y\times L_z=100a\times 200a$ and parameters in Hamiltonian are $M = 5$, $A = 50$, $k_W = 1.5$, $D_1 = 1$, and $D_2 = 4$.
      }
    \label{Fig2}
  \end{figure}
  \textit{ Edge states along $x$-$y$ plane.---}
Once the Weyl semimetal slab is finite in the $z$ direction, there exist edge states along $x$-$y$ plane [see Fig.~\ref{Fig1}(b)].
First, we study the edge states under a perpendicular magnetic field, i.e., $B_x=B_z=0$.
 On the top surface, electrons move along the Fermi arc from ``$-$'' Weyl node to ``$+$'' Weyl node to form a semi-circle as depicted by the red curve in Fig.~\ref{Fig1}(b), and transit into the bulk chiral Landau levels. In the bulk, electrons propagate to the bottom surface along a straight line in the $\bm{B}$ direction [see the dashed blue lines in Fig.~\ref{Fig1}(b)]. Since $\dot{\bm{k}}=0$, electrons stay at ``$+$'' Weyl node across the bulk. Due to the block of the back side surface, the cyclotron motion is forbidden on the bottom surface~(Sec. S2 of \cite{SM}). They have to bounce back to the bulk, however, scatter into the chiral Landau levels at ``$-$'' Weyl node. Since the Fermi velocities of the chiral Landau levels at different weyl nodes are opposite, electrons propagate back along $\bm{B}$ to the top surface [see the dashed green lines in Fig.~\ref{Fig1}(b)]. The electrons stay at ``$-$'' Weyl node before return to the Fermi arc on the top surface. Then, the above semiclassical motion is repeated. Eventually, they form conducting channels along the $x$ direction near the edge of the top surface. Similarly, the opposite conducting channels are formed by the same logic near the opposite edge of the bottom surface.

Second, it is necessary to generalize the above picture to the cases of a tilted magnetic field. The tilting angle $\theta$ refers to the angle $\bm{B}$ from the y axis, i.e., $\theta=\arctan(B_z/B_y)$ with $B_x=0$.  According to Eq.~\ref{eq2}, only $B_y$ drives electrons to move along the Fermi arcs, so the cyclotron motion on the surfaces will not be affected. For $\theta>0$, once the electrons reach the ``$+$'' Weyl node on the  top surface, they again transit into the chiral Landau level and will propagate parallel to $\bm{B}$. Unlike the $\theta=0$ case, the electrons will reach the back surface before arriving at the bottom surface [see the dashed blue lines in Fig.~\ref{Fig1}(c)]. Then, they will bounce back, scatter from ``$+$'' node to ``$-$'' node in $\bm{k}$ space and return to the top surface [see the dashed green lines in Fig.~\ref{Fig1}(c)]. Similar to the $\theta=0$ case, they also form the conducting channels.

Astonishingly, a magnetic field with $\theta<0$ can even shift the location of conducting channels [see Fig.~\ref{Fig1}(d)]. The semi-circle trajectory on the top surface makes the conducting channel tend to approach the back surface, while trajectory induced by the bulk chiral Landau modes have opposite effect. A small $\theta$ will keep the location of edge states similar as the $\theta>0$ case. However, when $\theta$ exceeds a critical value $\theta_c$, the latter mechanism dominates, and electrons from the top surface will hit the front surface rather than the back one ~(Sec. S2 of \cite{SM}). Consequently, the edge states will be spatially shifted [Fig.~\ref{Fig1}(d)].

For an infinite Weyl semimetal along $x$ direction, the above picture predicts two key signatures of LDOS on its $y$-$z$ cross section. Firstly, the edge states are distributed near the two diagonal corners of the cross section for $\theta=0$ and $\theta>0$ [see Fig.~\ref{Fig1}(b)(c)]. Contrarily, for $\theta<\theta_c<0$, the edge states will be shifted to the other two diagonal corners [see Fig.~\ref{Fig1}(d)]. Secondly, the bulk chiral Landau levels can also be observed and their LDOS is parallel to $\bm{B}$. 

We numerically calculate the corresponding LDOS in Fig.~\ref{Fig2}. It shows a good agreement with our semiclassical picture. For example,
for $\theta=15^\circ$ and $0^\circ$, the LDOS mainly concentrates around the hinges between the top (bottom) and the back (front) surfaces, indicating the diagonal distribution of edge states [Fig.~\ref{Fig2}(a)-(d)]. While for $\theta=-15^\circ<\theta_c$, the spatial shift of the edge states is demonstrated by the LDOS mainly concentrating around the hinges between the top (bottom) and the front (back) surfaces [Fig.~\ref{Fig2}(e)-(f)]. Further, compared to a perpendicular $B$, the LDOS is highlighted by the wedge shape (with direction parallel to $\bm{B}$) under a tilted $\bm{B}$. This feature is explained by the involvement of the bulk chiral Landau states and their orientations being consist to $\theta$.

\begin{figure}[b!]
    \includegraphics[width=\columnwidth]{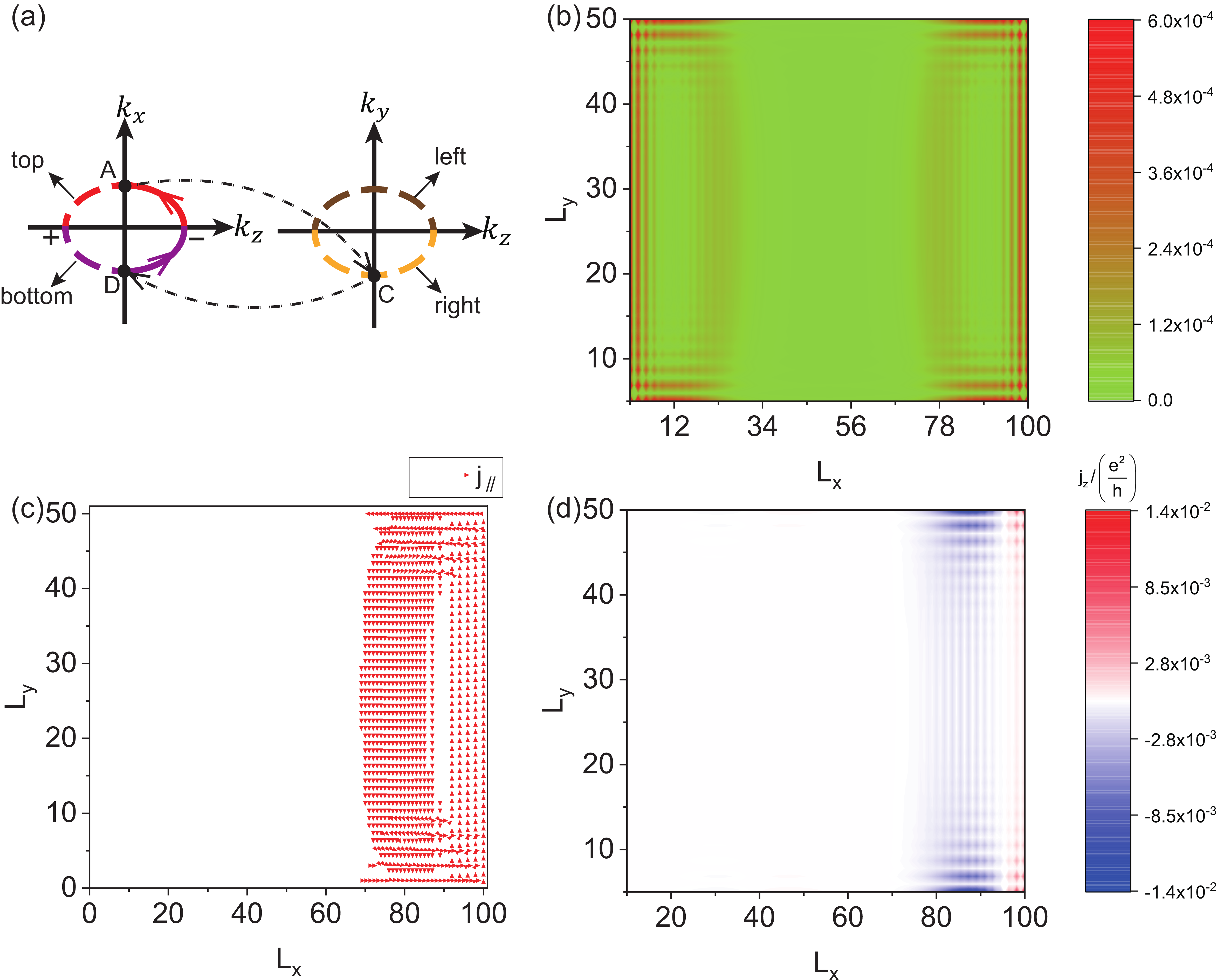}
    \caption{(a) A shematic diagram for the edge states near the right surface in the momentum space. (b) LDOS on the $x$-$y$ cross section of an infinite Weyl semimetal along the $z$ direction. (c) In-plane component of the local current density in the $x$-$y$ plane. (d) The $z$ component of the local current density. The current is injected from the positive $z$ direction, so only the right part of the edge modes are excited. The cross section size of the device is $L_x=100a$ and $L_y=50a$.
      }
    \label{Fig3}
  \end{figure}

  \textit{ Edge states along $y$-$z$ plane.---}
Once the Weyl semimetal is confined in the $x$ direction, the edge states will appear near the left and right surfaces [Fig.~\ref{Fig1}(b)]. The motion of an electron in momentum space is schetched in Fig.~\ref{Fig3}(a). When electrons on the top surface move from ``$-$'' Weyl node to $A$ point [red line in Fig.~\ref{Fig3}(a)], they undergo a solid red trajectory and then encounter the right surface in real space [see the red lines in Figure~\ref{Fig1}(b)]. At the hinge, points $A$ and $C$ are equivalent, due to shared $k_z$ between the top and right surfaces. For the topologically nontrivial right surface, $C$ point occupied by electrons is on the Fermi arc. During the motion on the right surface, the in-plane component of $\dot{\bm{k}}$ is zero according to Eq. (2). Therefore, they stay at $C$ all the time and acquire a velocity to reach the bottom surface [see the orange lines in Fig.~\ref{Fig1}(b)]. Meanwhile, points $C$ and $D$ are equivalent as well, because of the shared $k_z$ between the bottom and right surfaces. On the bottom surface, electrons move from $D$ point to ``$-$'' Weyl node (the solid purple line in Fig.~\ref{Fig3}(a)), and correspondingly undergo a purple trajectory in the real space. Then, they enter the bulk and travel to the top surface via the bulk chiral Landau levels (green lines in Fig.~\ref{Fig1}(b)). Therefore, the edge states near the right surface are formed by repeating the above motion. The edge states near the left surface can be understood in the same way. Here, we emphasize that both the chiral Landau and Fermi arc states in the $y$-$z$ surfaces are involved to form the edge states, and this is beyond the conventional Weyl orbit theory ~\cite{acpotterNC,zhangyiSR}.

We demonstrate the above picture by investigating the LDOS and the local current density with an infinite Weyl semimetal along the $z$ direction. The edge states can be captured by special features of these two physical quantities in the $x$-$y$ cross section. Because of the side surface states and bulk chiral Landau levels connecting the top and bottom surfaces [Fig.~\ref{Fig1}(b)], there should be handle-shaped LDOS in the $x$-$y$ cross section. This prediction is consistent with the LDOS calculation in Fig.~\ref{Fig3}(b), where the local maxima of LDOS around $x=0~(100)$ and $x=22~(78)$, separately correspond to the Fermi-arc states and the chiral Landau level states. Further, the fact that the bulk chiral Landau level states are more extended than the Ferm arc states on the side surfaces are also observed. Fig.~\ref{Fig3}(c) plot the in-plane component of the local current density. A counter-clockwise handle-shaped loop is observed, in agreement with the edge states along the $y$-$z$ plane, where electrons go clockwise in the $x$-$y$ plane while propagating along the $z$ direction.
Moreover, the propagation along the $z$ direction is originated from both top and bottom surface states as shown in Fig.~\ref{Fig1}(b). Indeed, the $z$ component of the local current density in Fig.~\ref{Fig3}(d) concentrates around $y=0$ and $y=L_y$.

  \begin{figure}[t!]
    \includegraphics[width=\columnwidth]{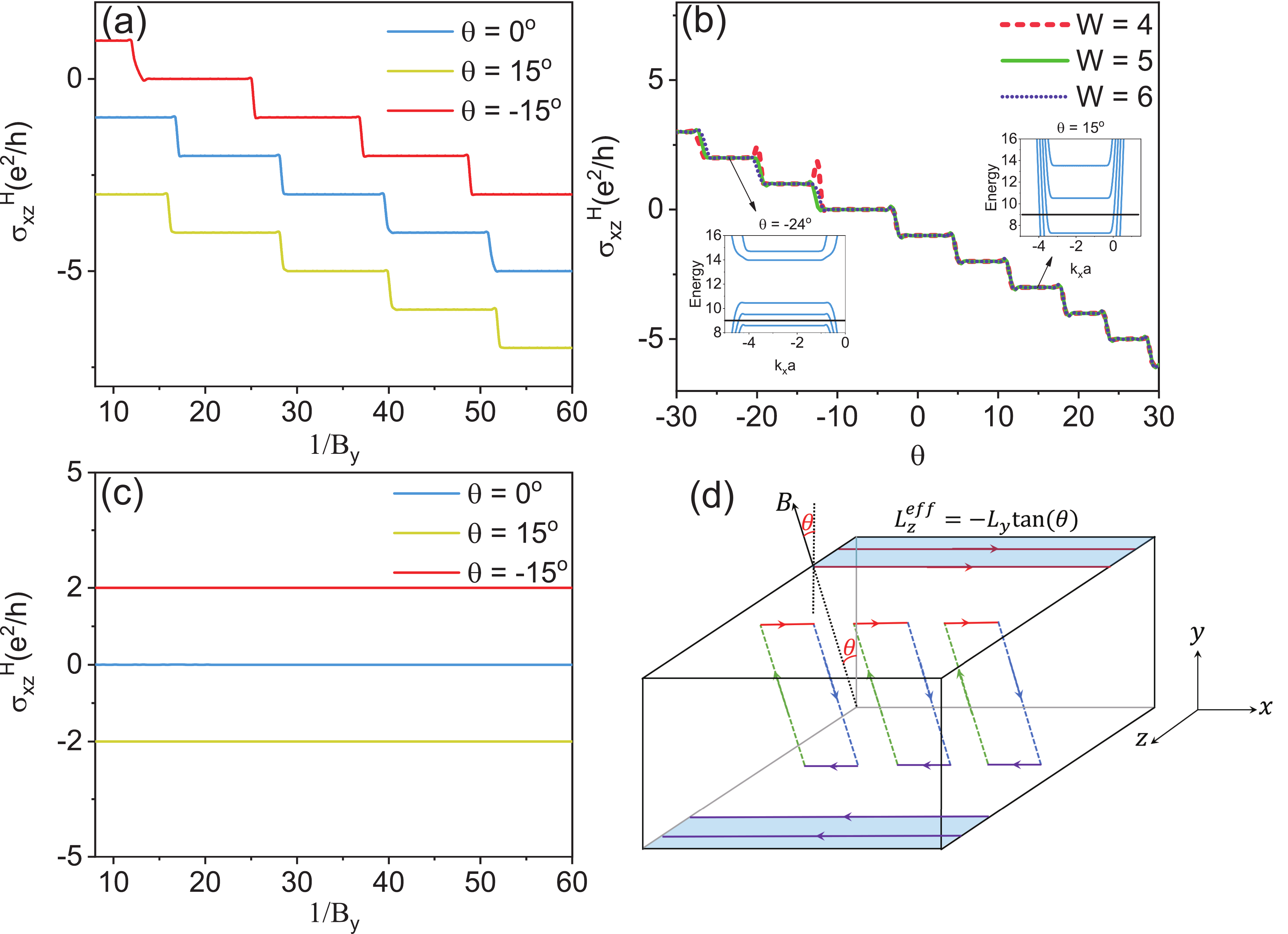}
    \caption{The Hall conductance $\sigma_{xz}^H$ at $E_F=E_0$ of a four-terminal device. (a) The dependence of $\sigma_{xz}^H$ on $1/B_y$ under angle $\theta=-15^\circ,0^\circ,15^\circ$. (b) The $\theta$ dependence of $\sigma_{xz}^H$ with a fixed $B_y=1/10$. The left and right insets show energy-bands at $\theta=-24^\circ$ and $15^\circ$.
    (c) $\sigma_{xz}^H$ with $D_1=D_2=0$, i.e. the Fermi surface area enclosed by the Fermi arcs $S_{\bm{k}}=0$, revealing the net contribution from the chiral bulk Landau levels. (d) The schematic diagram of the effective area supporting edge states under a tilted magnetic field. The parameters in numerical calculation are $D_1 = 1$, $D_2 = 4$ for (a)(b) and $D_1 = D_2=0$ for (c)(d). The cross section size is $L_z=40a$ and $L_y=20a$.
      }\label{Fig4}
  \end{figure}

  \textit{The Hall transport.---}
  Since the edge states can be manipulated by the chiral Landau levels under different magnetic field directions, the Hall conductance is expected to depend on the tilting angle $\theta$. We study the Hall conductance of a four-terminal device by the NEGF method numerically~\cite{SM}, and it is defined as
  $\sigma^H_{xz}=I_{x}/V_{xz}$,
  where $I_{x}$ is the longitudinal current and $V_{xz}$ is the transverse voltage difference. Interestingly, we find $\sigma_{xz}^H$ does not depend on the location of the transverse leads, consistent with the existence of the bulk chiral Landau levels [Fig.~\ref{Fig2}]. To examine the robustness of the Hall plateaus, Anderson-type disorder is introduced in the model~\cite{jianghuaPRB}. The numerical results are shown in Fig.~\ref{Fig4}.

  Figure~\ref{Fig4}(a) shows the $1/B_y$ dependence of $\sigma_{xz}^H$ with a fixed tilting angle. $\sigma_{xz}^H$ under different tilting angles with a fixed $B_y$ verifies the manipulation by the chiral Landau levels and indicates tunable initial values. Especially for $\theta=-15^\circ<\theta_c$, $\sigma_{xz}^H$ goes down from one to zero and becomes negative. Here, we do not reverse the magnetic field, but do change the sign of $\sigma_{xz}^H$, which is distinctly different from 2D QHE. Subsequently, we simulate the $\theta$ dependence of $\sigma_{xz}^H$ under a given $B_y$. Fig.~\ref{Fig4}(b) shows well quantized plateaus of $\sigma_{xz}^H$ under different disorder strengths. Owing to the fixed $B_y$, the contribution to $\sigma_{xz}^H$ from gauge potential is fixed. Thus, the chiral bulk Landau levels indeed manipulate the edge states [It's also confirmed by the number of edge states under different $\theta$ in Fig.~\ref{Fig4}(b)]. Moreover, this phenomenon can experimentally rule out the possibility of the QHE formed by 2D electron gas in Weyl semimetals.

  Combining the above semiclassical picture and quantization of Landau levels~\cite{luhaihzouPRL}, we find 3D QHE in Weyl semimetals depends on both surface states and bulk chiral Landau levels, and can be written as~(Sec.~S5 of \cite{SM})
  \begin{equation}
    \sigma_{xz}^H=\frac{e^2}{h}(n_0+n),~~~n=0,-1,-2,...\label{eq4}
  \end{equation}
  where $n$ is originated from the Fermi surface area $S_{\bm{k}}$ enclosed by the Fermi arcs and approximately proportional to $-\hbar S_{\bm{k}}/(2\pi eB_y)$~\cite{luhaihzouPRL}. Besides, we find there is an additional $n_0$ originated from the bulk chiral Landau levels and dependent on the direction of $\bm{B}$.

  To figure out the initial value $n_0$, we set $S_{\bm{k}}=0$ to exclude the effect of the surface states with $n=0$. In Fig.~\ref{Fig4}(c), the fact that $\sigma_{xz}^H$ is odd with the tilting angle and independent of $B_y$ confirms the existence of $n_0$ in Eq.~\ref{eq4}.
 We find $n_0$ can be understood by the intrinsic QAH states in Weyl semimetals~(Sec. S3 of \cite{SM}). In the absence of a magnetic field, the Fermi-arc states are indeed edge states surrounding the $z$ axis~\cite{jianghuaPRB02,reviewRMP}. A perpendicular magnetic field can deform all edge states into localized Weyl orbits and lead to zero Hall conductance ($n_0=0$). For a tilted magnetic field, not all edge states are localized, because of the tilted chiral Landau levels. As illustrated in Fig.~\ref{Fig4}(d), the blue surface accommodates the edge states and the rest accommodates the localized Weyl-orbit states. It introduces an effective width, $L_z^{(eff)}=-L_y\operatorname{tan}\theta$. Quantitatively, $n_0$ in Eq.~\ref{eq4} is exactly the number of edge states in blue surface~\cite{roundingdown},
  \begin{equation}
    n_0=\left\lfloor\frac{2k_w}{\frac{2\pi}{L_z^{(eff)}}}\right\rfloor=-\left\lfloor \frac{k_w}{\pi}L_y\operatorname{tan}\theta\right\rfloor.
  \end{equation}
It explains why $\sigma_{xz}^H$ can be tuned by the tilting angle. For details, see supplemental materials~\cite{SM}.

Recently, $\rm MnBi_2Te_4$ was proposed to be a magnetic Weyl semimetal where only one pair of Weyl nodes exists~\cite{xuyongPRB,wangjianhighchernnumber}. Higher-Chern-number QAH in $\rm MnBi_2Te_4$ was also successfully observed~\cite{wangjianhighchernnumber}, and Fermi-arc states can exist on side surfaces. We can rotate the sample by $90^\circ$ and expect it to be an ideal platform to realize our distinctive Hall phenomena. Another candidate for experiments is the newly found magnetic Weyl semimetal $\rm Co_3Sn_2S_2$~\cite{SCIENCE201901,SCIENCE201902,liuenkeSCIENCE}. For an extremely large magnetic field or in the quantum limit, the observed Hall conductance is dominated by $n_0$. Sign change of $\sigma_{xz}^H$ ought to be observed when $\theta$ exceeds $\theta_c$ under a fixed $B_y$, and this will confirm the role of the chiral bulk Landau levels in our theory.
Additionally, the LDOS of the top or bottom surfaces measured by scanning tunneling microscopy can reflect the distribution of the edge states. For instance, as illustrated in Fig.~\ref{Fig2}, the spatial shift of the edge states in LDOS can be measured by tilting $\theta$ around $\theta_c$.

  \textit{Conclusion.---}We find a global picture to describe the edge states of 3D QHE in Weyl semimetals. The bulk chiral Landau levels parallel to the magnetic field encode the quantization of edge states.
  Therefore, a tilted magnetic field can control the edge states and lead to distinctive Hall transport phenomena.
  Our work provides a comprehensive understanding to the topological nature of 3D QHE in Weyl semimetals.

  \textit{Acknowledgement.---} We thank Chui-Zhen Chen and Qing-Feng Sun for fruitful discussion. This work is financially supported by the National Basic Research Program of China (Grants No. 2017YFA0303301, and No. 2019YFA0308403) and the National Natural Science Foundation of China (Grants No. 11534001, No. 11674028, and No. 11822407).

%

  \begin{widetext}
    \pagebreak
    \newpage
    \setcounter{equation}{0}
    \setcounter{figure}{0}
    \setcounter{table}{0}
    \setcounter{page}{0}
    \makeatletter
    \renewcommand{\thepage}{S\arabic{page}} 
    \renewcommand{\thesection}{S\arabic{section}}  
    \renewcommand{\thetable}{S\arabic{table}}  
    \renewcommand{\thefigure}{\arabic{figure}}
    \renewcommand\figurename{Figure S}

  \begin{center}
  \textbf{Supplementary Materials for ``3D Quantum Hall Effect Manipulated by Chiral Landau Levels in Weyl Semimetals''}
  \end{center}
  \begin{center}
    Hailong Li$^{1}$, Haiwen Liu$^{2}$, Hua Jiang$^{3,4,*}$, and X. C. Xie$^{1,5,6,\dag}$
    \end{center}
  
    \begin{center}
      $^1$~International Center for Quantum Materials, School of Physics, Peking University, Beijing 100871, China
      
      $^2$~Center for Advanced Quantum Studies, Department of Physics, Beijing Normal University, Beijing 100875, China
      
      $^3$~School of Physical Science and Technology, Soochow University, Suzhou 215006, China
      
      $^4$~Institute for Advanced Study, Soochow University, Suzhou 215006, China
      
      $^5$~Beijing Academy of Quantum Information Sciences, Beijing 100193, China
      
      $^6$~CAS Center for Excellence in Topological Quantum Computation, University of Chinese Academy of Sciences, Beijing 100190, China
      \end{center}
      \tableofcontents
\section{S1. Semiclassical motion of electrons}
\begin{figure}[htbp]
  \centering
  \includegraphics[width=0.45\columnwidth]{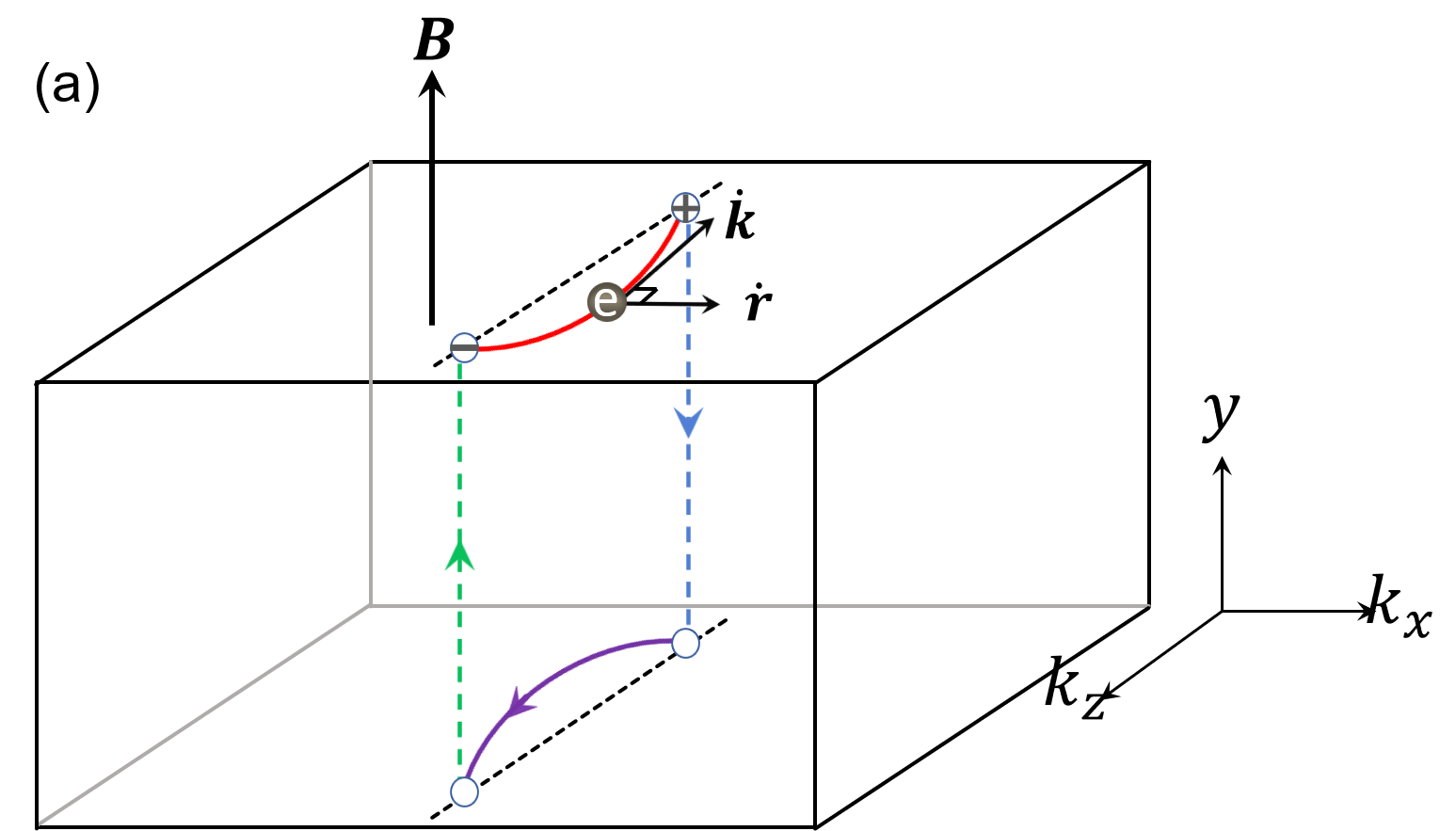}
  \includegraphics[width=0.45\columnwidth]{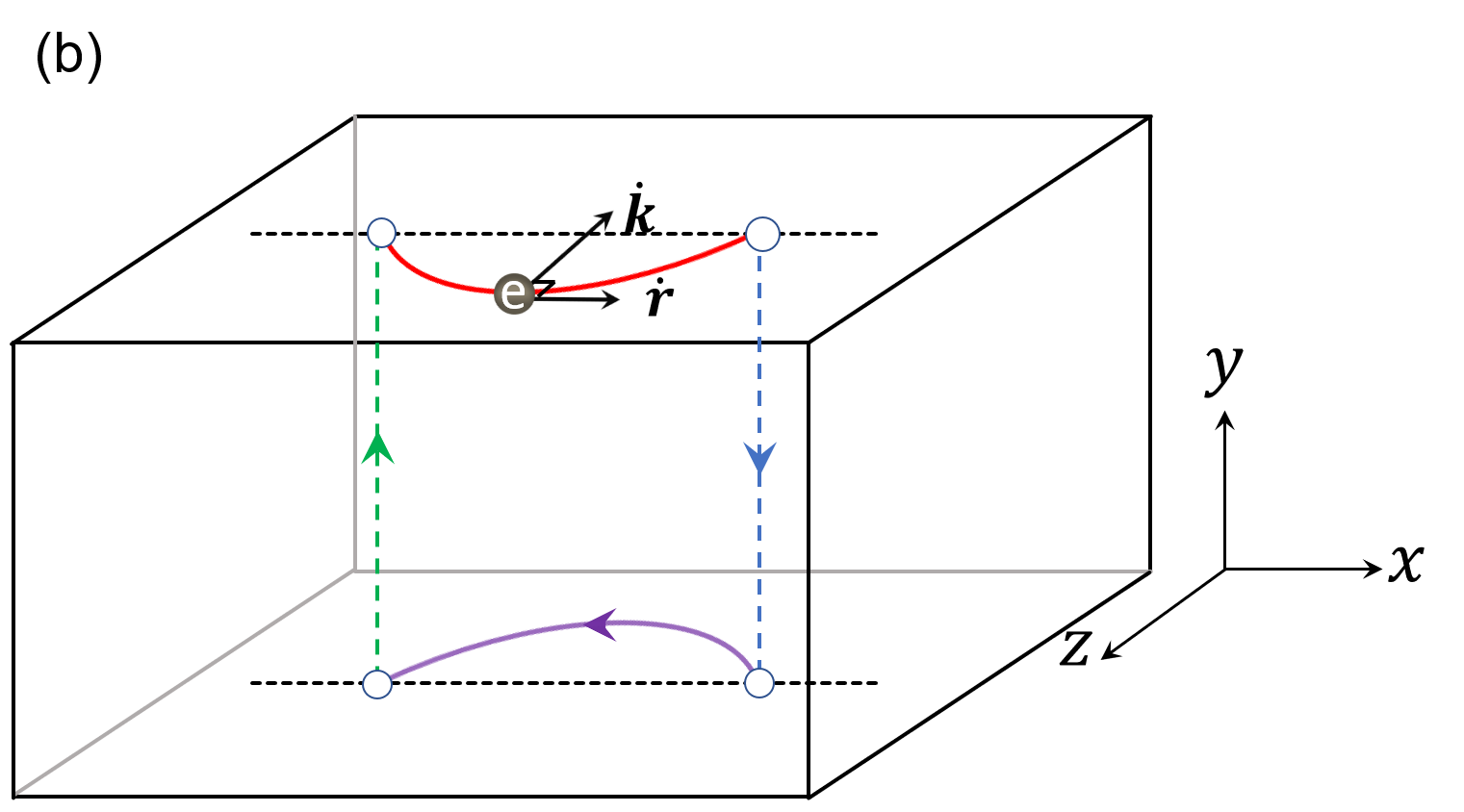}
  \caption{A Weyl semimetal slab under an external magnetic field. (a) Semiclassical motion of electrons is depicted in hybrid real space ($y$) and momentum space ($k_x,k_z$). Green and blue lines are from Chiral bulk Landau levels at Weyl nodes. (b) It shows the corresponding trajectory in the real space.}\label{semi_motion}
  \end{figure}
  Fig.S~\ref{semi_motion} shows semiclassical motion derived from Eq.~2 in the main text, which establishes a map between the momentum ${\boldsymbol{k}}$ and the position ${\boldsymbol{r}}$.
  Electrons on the topological surfaces are from the Fermi arcs. Therefore, $\dot{\bm{r}}$ of the electrons is orientated along the normal of Fermi arcs and $\dot{\bm{k}}$ is tangent to the arcs, i.e. $\dot{\bm{k}}\perp \dot{\bm{r}}$. In $\boldsymbol{k}$ space, electrons slide along Fermi arcs and map a trajectory into the real space. As for the electrons in the bulk or trivial surfaces, they are from the chiral Landau levels. Thus, it makes $\dot{\bm{k}}$ zero that $\dot{\bm{r}}$ is parallel to $\bm{B}$. Trajectories in real space are straight lines parallel to $\bm{B}$.
\section{S2. Critical angle for Hall conductance}
\begin{figure}[htbp]
  \centering
  \includegraphics[width=0.45\columnwidth]{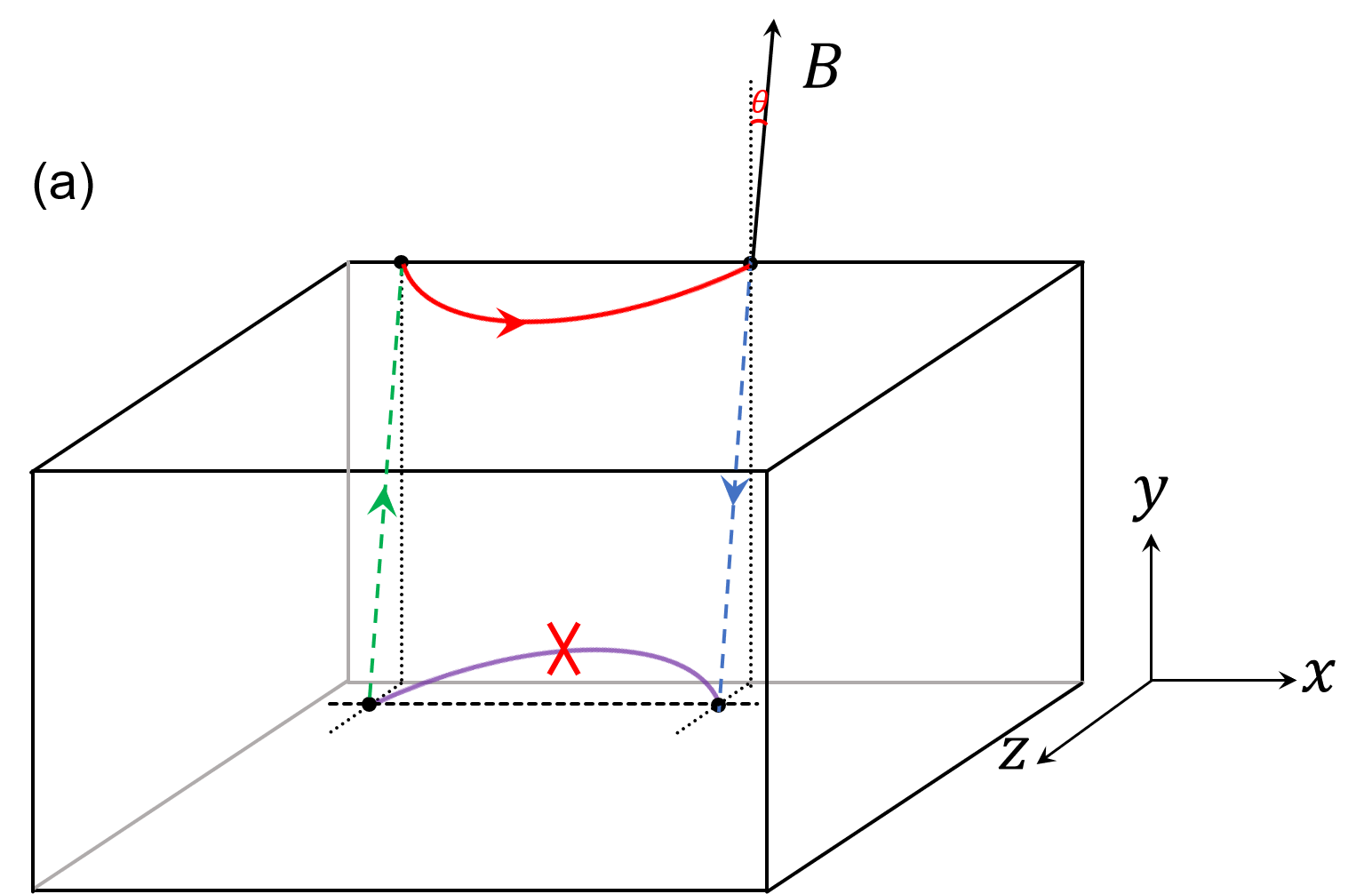}
  \includegraphics[width=0.45\columnwidth]{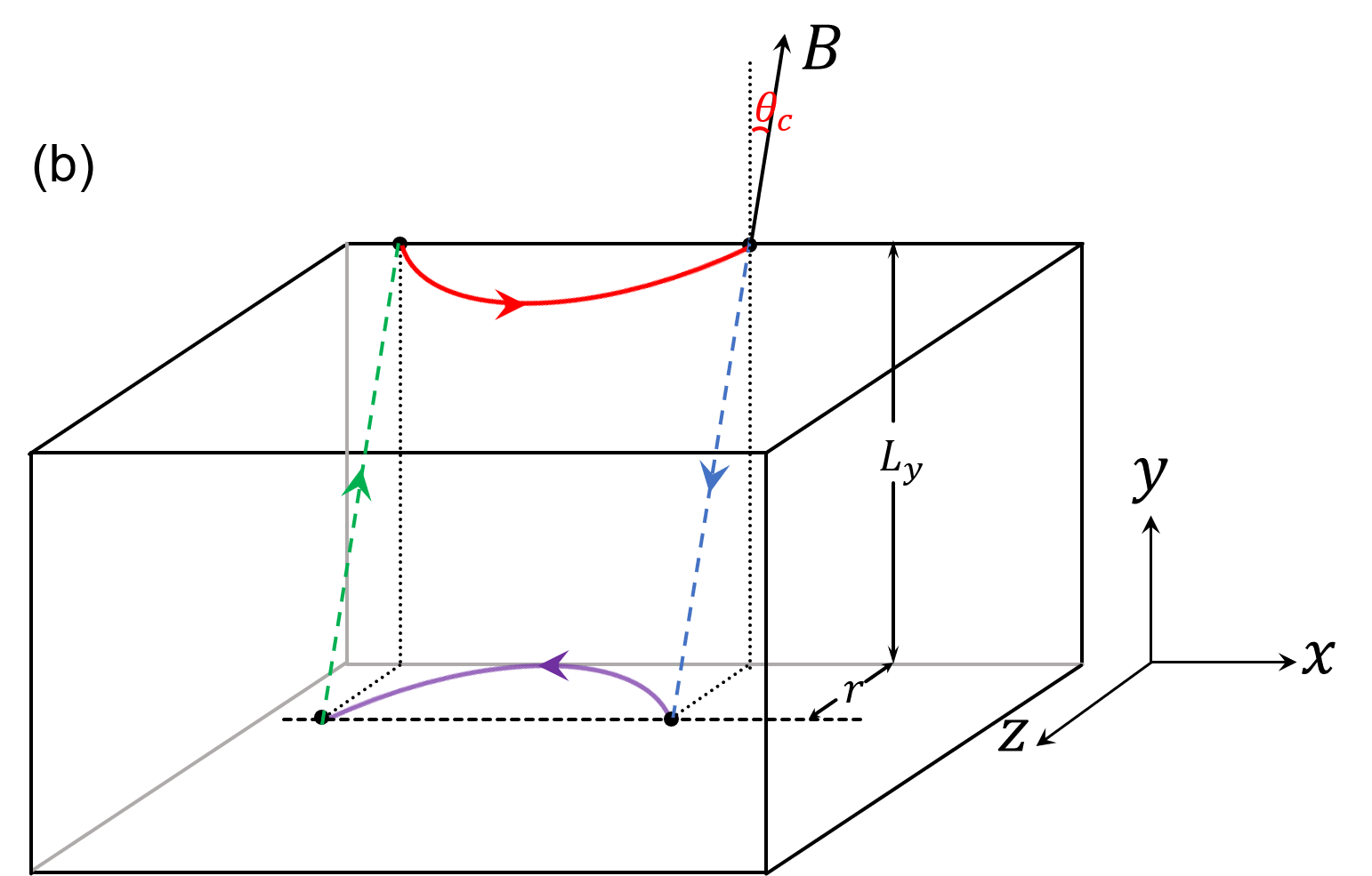}
  \caption{(a) The basic unit of the skipping orbit near the back surface for a small tilting angle. (b) It shows the critical angle $\theta_c$ to form a complete Weyl orbit near the back surface.}\label{critical_angle}
  \end{figure}
In the Fig.~4 of the main text, we find that the Hall conductance, $\sigma_{xz}^H$, changes its sign at a critical tilting angle $\theta_c$. Here, $\theta_c$ is a negative value rather than zero. The existence of $\theta_c$ can also be explained by our semiclassical picture as illustrated in Fig.S~\ref{critical_angle}. 

The semiclassical analysis begins with a basic unit of the skipping orbit near the back surface [Fig.S~\ref{critical_angle}(a)]. The electron on the top surface slides along the Fermi arc and then enters the bulk chiral Landau levels at ``$+$'' Weyl node. For a tilting angle, $\theta>\theta_c$, the position where the electron arrives at the bottom surface is so close to the back surface that the cyclotron motion at the bottom surface is not allowed. Thus, the electron has to bounce back to the top surface. It consequently forms a conducting channel which supports a negative Hall conductance in our calculations. 

That the critical angle exists means the complete Weyl orbit is allowed even if it touches the back surface when $\theta=\theta_c$. As illustrated in Fig.S~\ref{critical_angle}(b), the complete Weyl orbit makes the conducting state in Fig.S~\ref{critical_angle}(a) a localized state. Therefore, the conducting channel along the positive  $x$ direction changes to the vicinity of the front surface [see Fig.~1(d) in the main text]. Correspondingly, the Hall conductance becomes positive. What's more, $\theta_c$ can be estimated by the geometric relationships showed in Fig.S~\ref{critical_angle}(b):
\begin{equation}
  tan(\theta_c)=-\frac{r(B_y)}{L_y},
\end{equation}
where $r$ is the cyclotron radius depending on $B_y$ and $L_y$ is the thickness of the Weyl semimetal slab.

\section{S3. Hall conductance from bulk chiral Landau levels}
\begin{figure}[b!]
  \centering
  \includegraphics[width=\columnwidth]{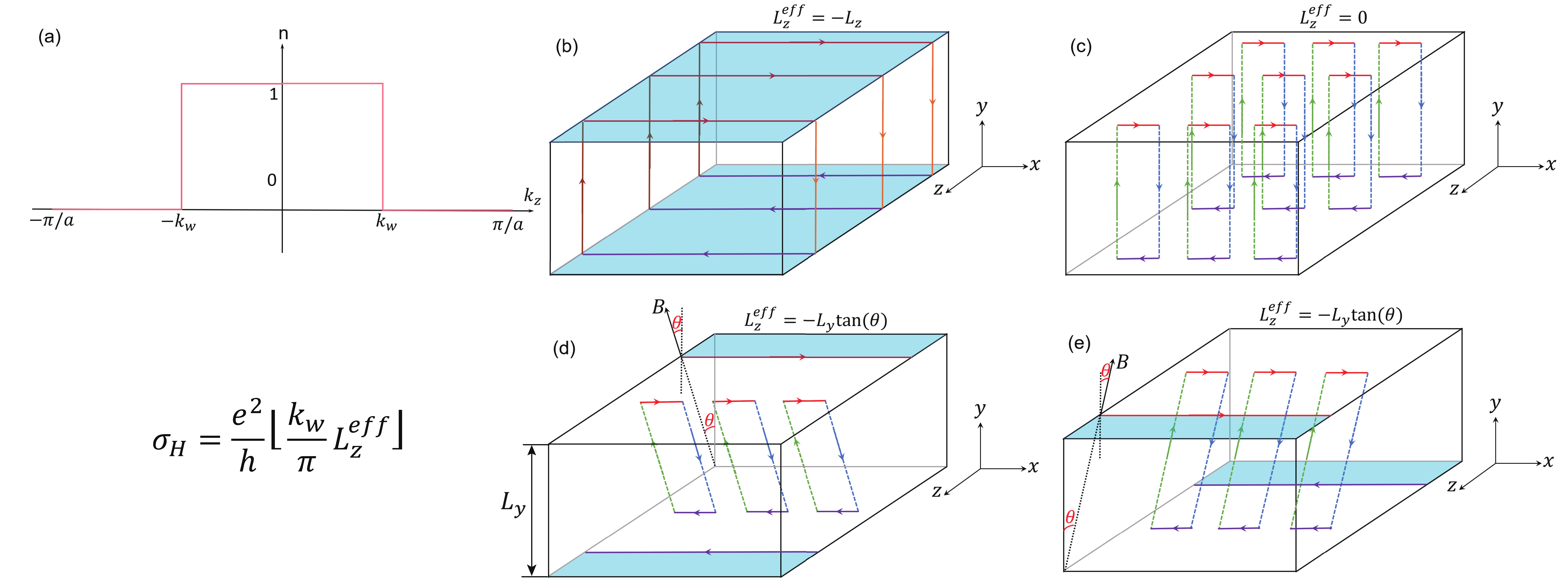}
  \caption{(a) It shows the Chern number as a function of $k_z$. (b) The topological surface states of a Weyl semimetal without the magnetic field. (c) Surface states are converted into localized states under a vertical magnetic field. (d) and (e) show the edge states under tilted magnetic fields. Here, red, purple, brown and orange lines are from Fermi arc states. Blue and green lines are from chiral bulk Landau levels. $\left\lfloor... \right\rfloor$ stands for rounding down.}
  \label{chernnumber}
\end{figure}

We will discuss the $n_0$ part of Hall conductance in detail. By setting $D_1=D_2$, we make $S_k$ zero and exclude the contribution from surface states to $\sigma_{xz}^H$, i.e., $\sigma_{xz}^H$ will not vary with the magnitude of the magnetic field. Thus,
\begin{equation}
  \sigma_{xz}^H=\frac{e^2}{h}n_0.
\end{equation}
$n_0$ can be estimated by the picture depicted in Fig.S~\ref{chernnumber}. For a 2D plane with fixed $k_z$, except for two Weyl nodes, it is well-defined to evaluate its Chern number. More explicitly in Fig.S~\ref{chernnumber}(a), $n=1$ for $-k_w<k_z<k_w$, while $n =0$ for $k_z<-k_w$ or $k_z>k_w$~\cite{SxugangPRL}. Each topologically nontrivial plane can be seen as a 2D quantum anomalous Hall insulator. Thus, a Weyl semimetal can be treated as a stack of 2D quantum anomalous Hall insulators with a total Chern number N,
\begin{equation}
  N=\left\lfloor \frac{2k_w}{\frac{2\pi}{L_z}}\right\rfloor=\left\lfloor\frac{k_w}{\pi}L_z\right\rfloor.\label{qah}
\end{equation}
Here, $L_z$ is the width along the direction of two Weyl nodes and $\left\lfloor... \right\rfloor $ stands for rounding down. Eq.~\ref{qah} reveals the cyclic surface states in Fig.S~\ref{chernnumber}(b). In the presence of a perpendicular magnetic field $B_y$, electrons on the top and bottom surfaces will form Weyl orbits. It means that the conducting surface states in Fig.S~\ref{chernnumber}(b) will deform into localized states in Fig.S~\ref{chernnumber}(c). The Weyl orbits in Fig.S~\ref{chernnumber}(c) look like rectangles for the zero $S_k$ and straight Fermi arcs. Moreover, these retangular Weyl orbits results in zero $\sigma_{xz}^H$ under perpendicular magnetic fields, unlike the curved ones in the main text. This result can be understood by our global picture in the main text. 

When the magnetic field is tilted, the quantum anomalous Hall states are not completely destroyed by the closed Weyl orbits. There remains some part of the top and bottom surfaces to host edge states. In Fig.S~\ref{chernnumber}(d) and S~\ref{chernnumber}(e), we use blue to mark the region occupied by edge states. The electrons outside the blue region form close Weyl orbits and become localized. However, the electrons inside the blue region can not tunnel to top or bottom surface to form complete Weyl orbits for the block of the front or back surface. The semiclassical trajectories of the edge states can be obtained by following the global pcture described in the main text. Because the one-dimensional chiral bulk Landau levels are parallel to the magnetic field, the blue region depends on the tilting angle $\theta$. According to the geometric relationships in Fig.S~\ref{chernnumber}(d) and S~\ref{chernnumber}(e), the effective width to support edge states is
\begin{equation}
  L_z^{(eff)}=-L_y\tan(\theta)\label{eq:n0}
\end{equation}
By the same logic of Eq.~\ref{qah}, the Hall conductance can be obtained directly,
\begin{equation}
  \sigma_{xz}^H=\frac{e^2}{h}\left\lfloor \frac{2k_w}{\frac{2\pi}{L_z^{(eff)}}}\right\rfloor=-\frac{e^2}{h}\left\lfloor\frac{k_w}{\pi}L_y\tan(\theta)\right\rfloor\label{n0}.
\end{equation}
Here, $\left\lfloor... \right\rfloor$ stands for rounding down. Moreover, when the magnetic filed is tilted from Fig.S~\ref{chernnumber}(d) to Fig.S~\ref{chernnumber}(e), it will cause the spatial shift of edge states and consequently change the sign of Hall conductance. Then, $n_0$ can be extracted from Eq.~\ref{n0},
\begin{equation}
  n_0=-\left\lfloor\frac{k_w}{\pi}L_y\tan(\theta)\right\rfloor.
\end{equation}
\section{S4. Edge states along side surfaces under tilted magnetic field}
\begin{figure}[htbp]
  \centering
  \includegraphics[width=0.49\columnwidth]{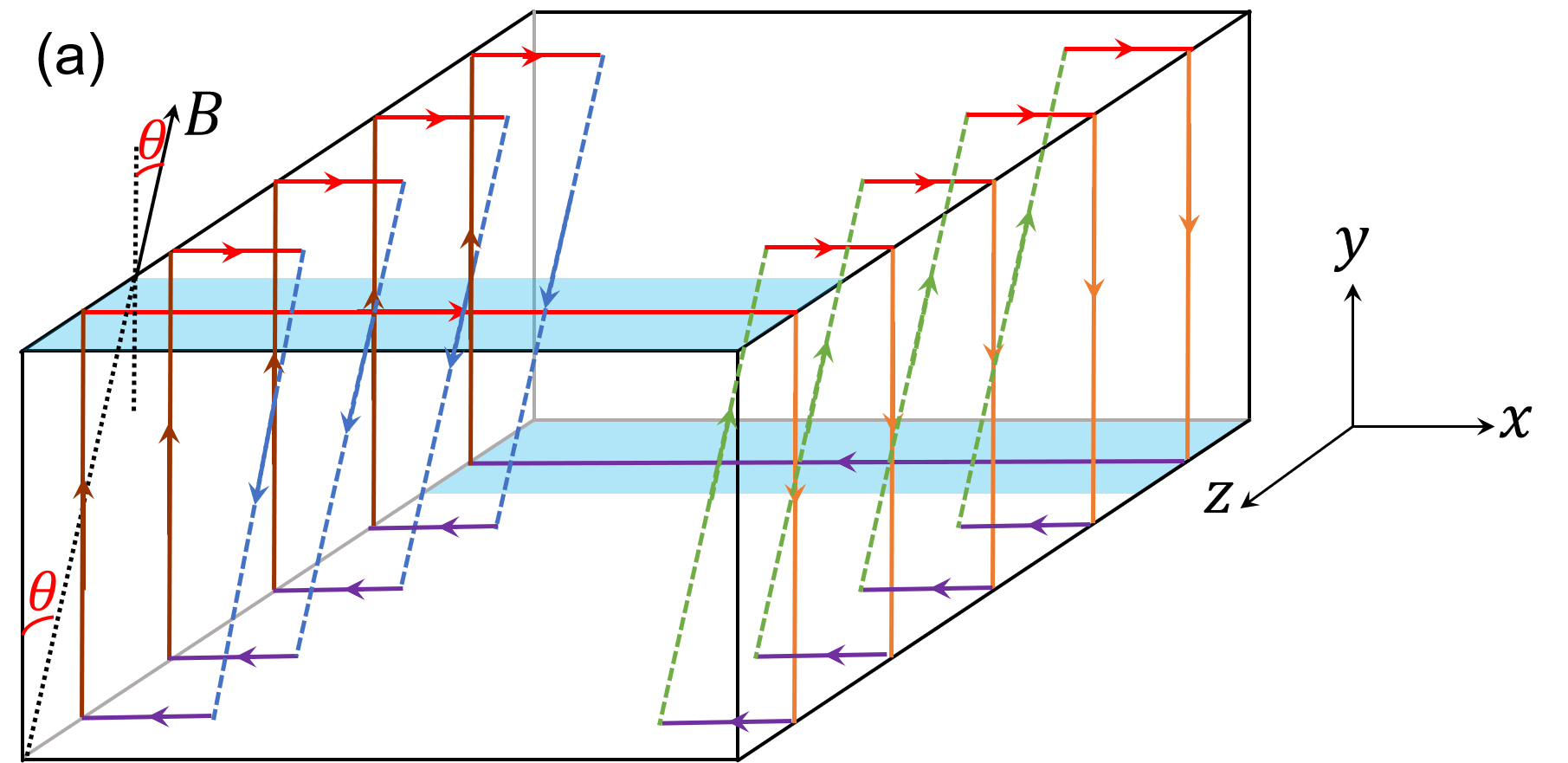}
  \includegraphics[width=0.49\columnwidth]{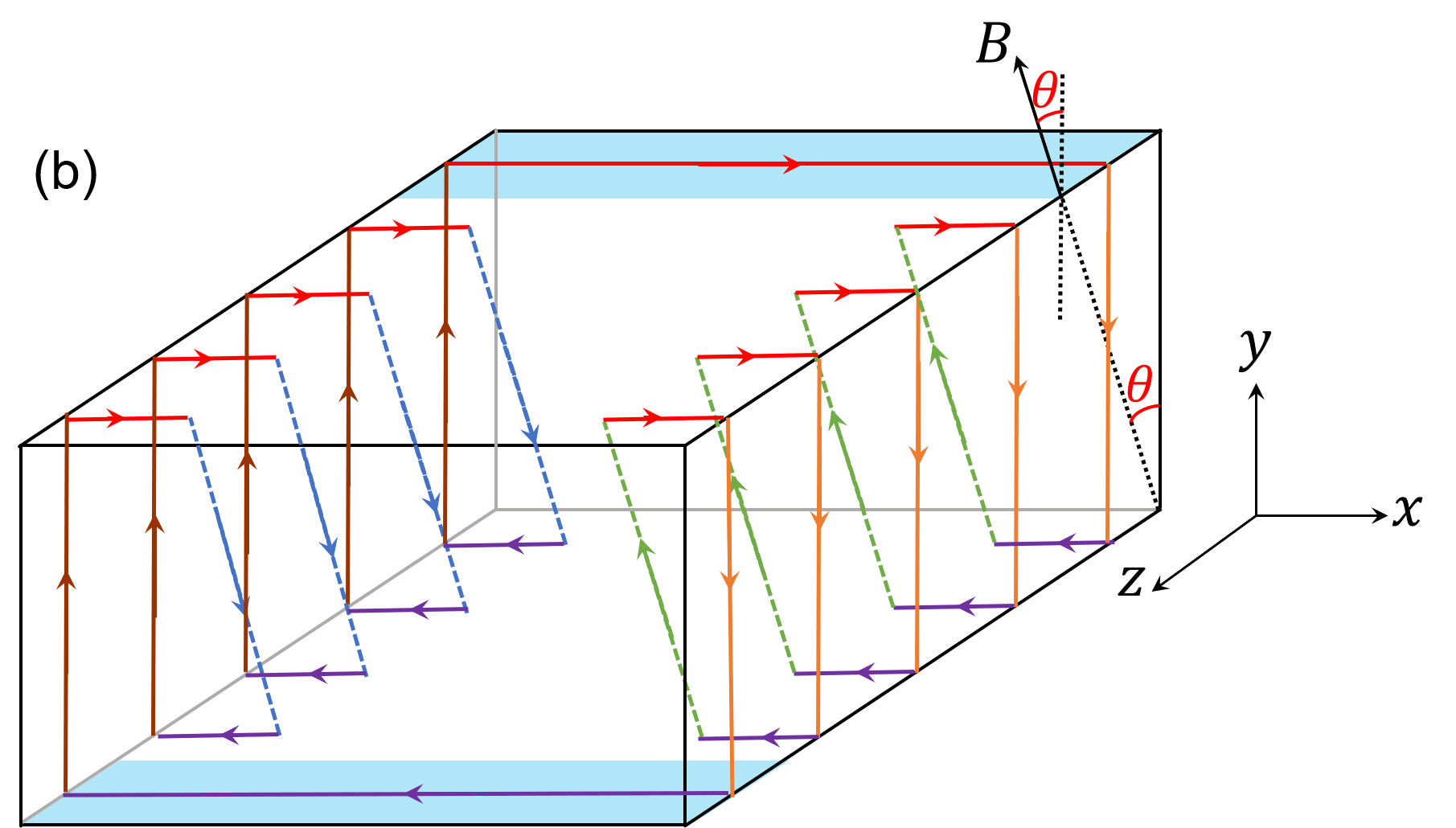}
  \caption{A semiclassical picture of edge states along $y$-$z$ plane under tilted magntic field. (a) $\theta<0$. (b) $\theta>0$. Here, we adopt $D_1=D_2=0$ for straight Fermi arcs. Red, purple, brown and orange lines are from Fermi-arc states. Blue and green lines are from chiral bulk Landau levels.}
  \label{side_surface}
\end{figure}
Without loss of generality, we set $D_1=D_2=0$ for straight Fermi arcs. The semiclassical picture of edge states along $y$-$z$ plane under tilted magnetic field is depicted in Fig.S~\ref{side_surface}. The edge states along $x$-$y$ plane 
can be obtained according to Fig.~1(c) and 1(d), and here, they are denoted by unidirectional lines for simplicity.

The motion of electrons in the momentum sapce is the same as Fig.~3(a) in the main text except for the curved Fermi arcs. Due to the straight Fermi arcs, the trajectories on the top and bottom surfaces are straight, too [see red lines in Fig.S~\ref{side_surface}]. The edge states on the top and bottom surfaces are connected by chiral bulk Landau levels parallel to $\bm{B}$ [see blue and green dashed lines in Fig.S~\ref{side_surface}] and side surface states [see brown and orange lines in Fig.S~\ref{side_surface}].
\section{S5. Schematic explanation to $\sigma_{xz}^H$}
\begin{figure}[htbp]
  \centering
  \includegraphics[width=0.3\columnwidth]{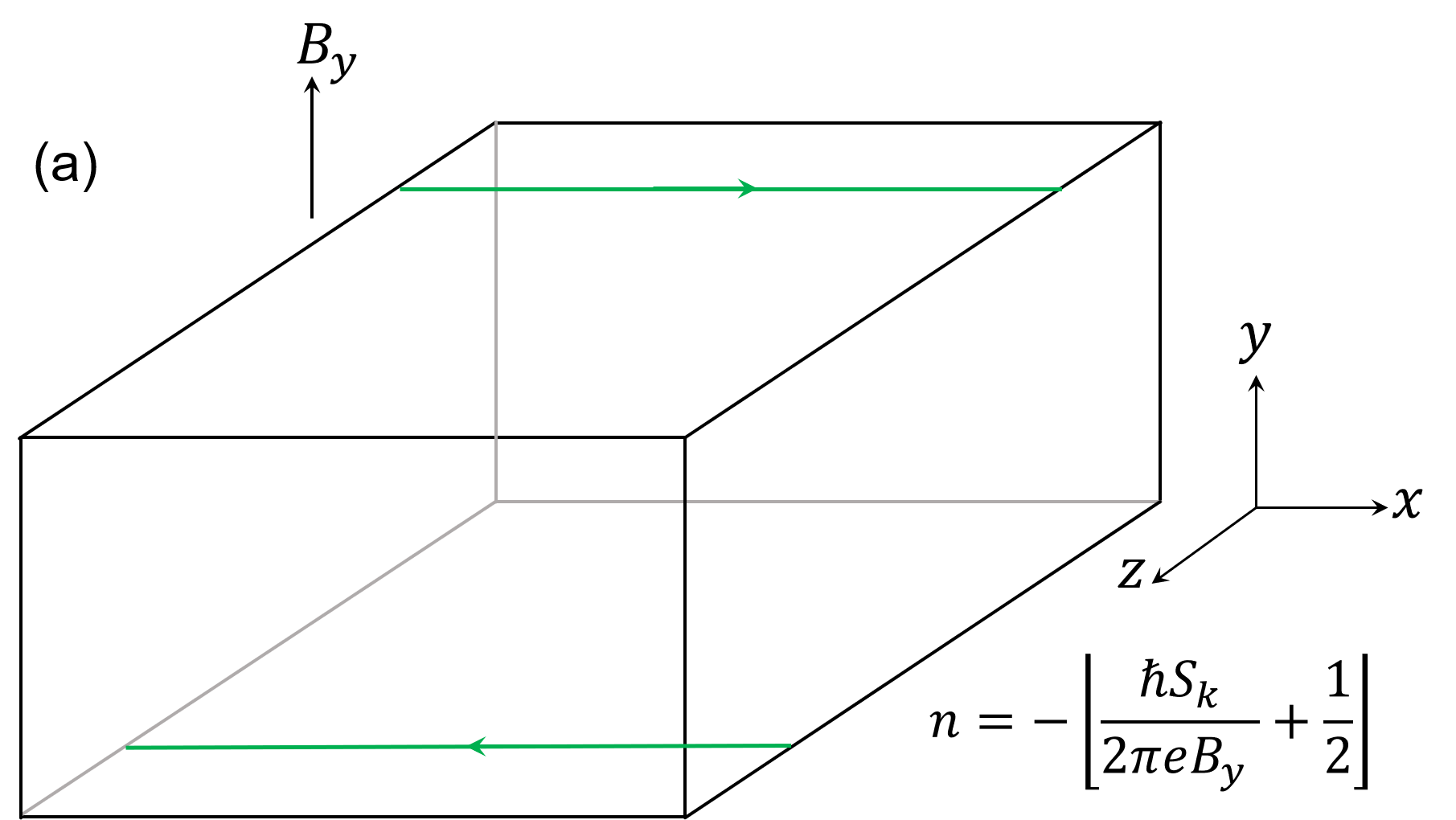}
  \includegraphics[width=0.3\columnwidth]{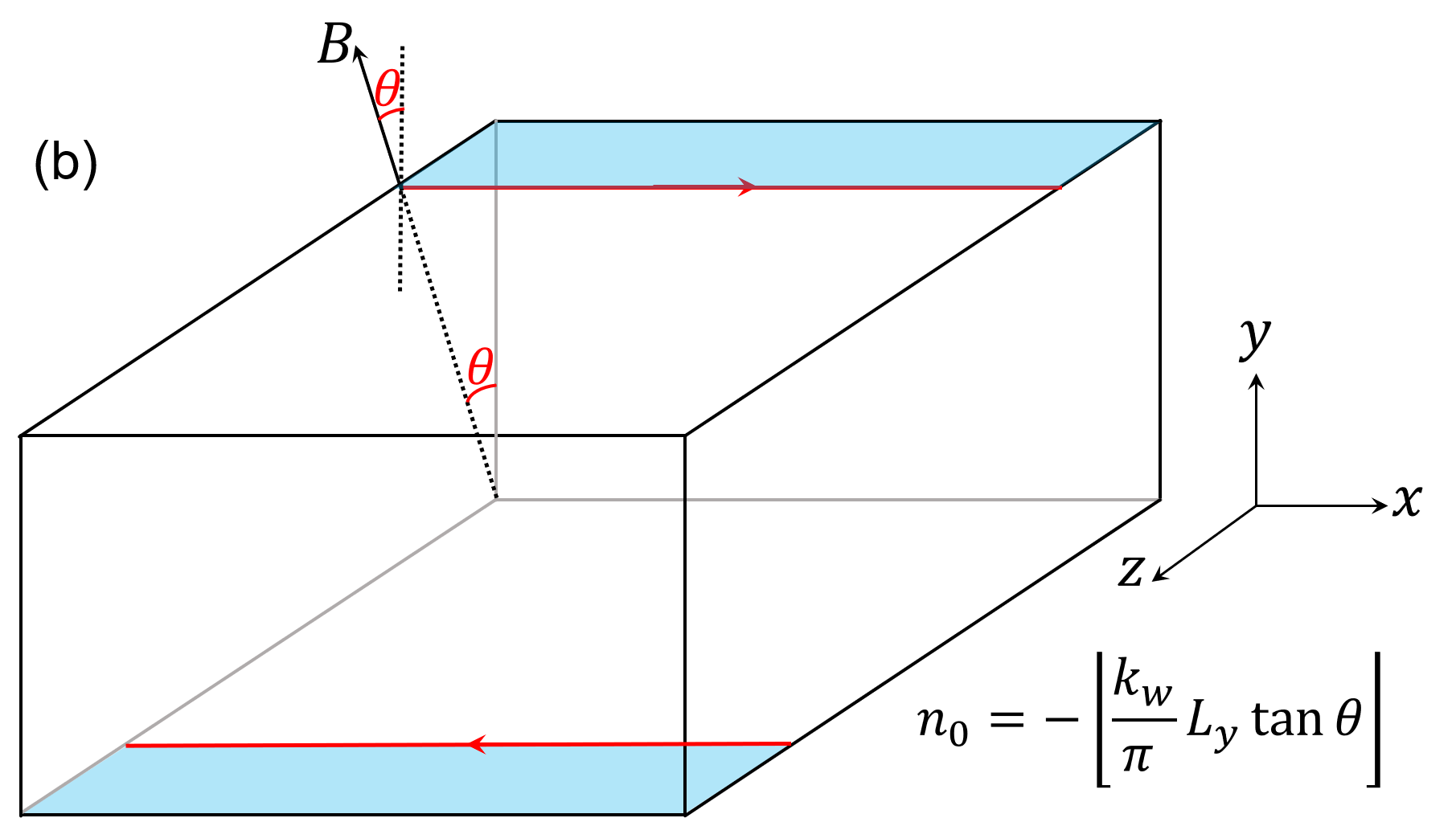}
  \includegraphics[width=0.3\columnwidth]{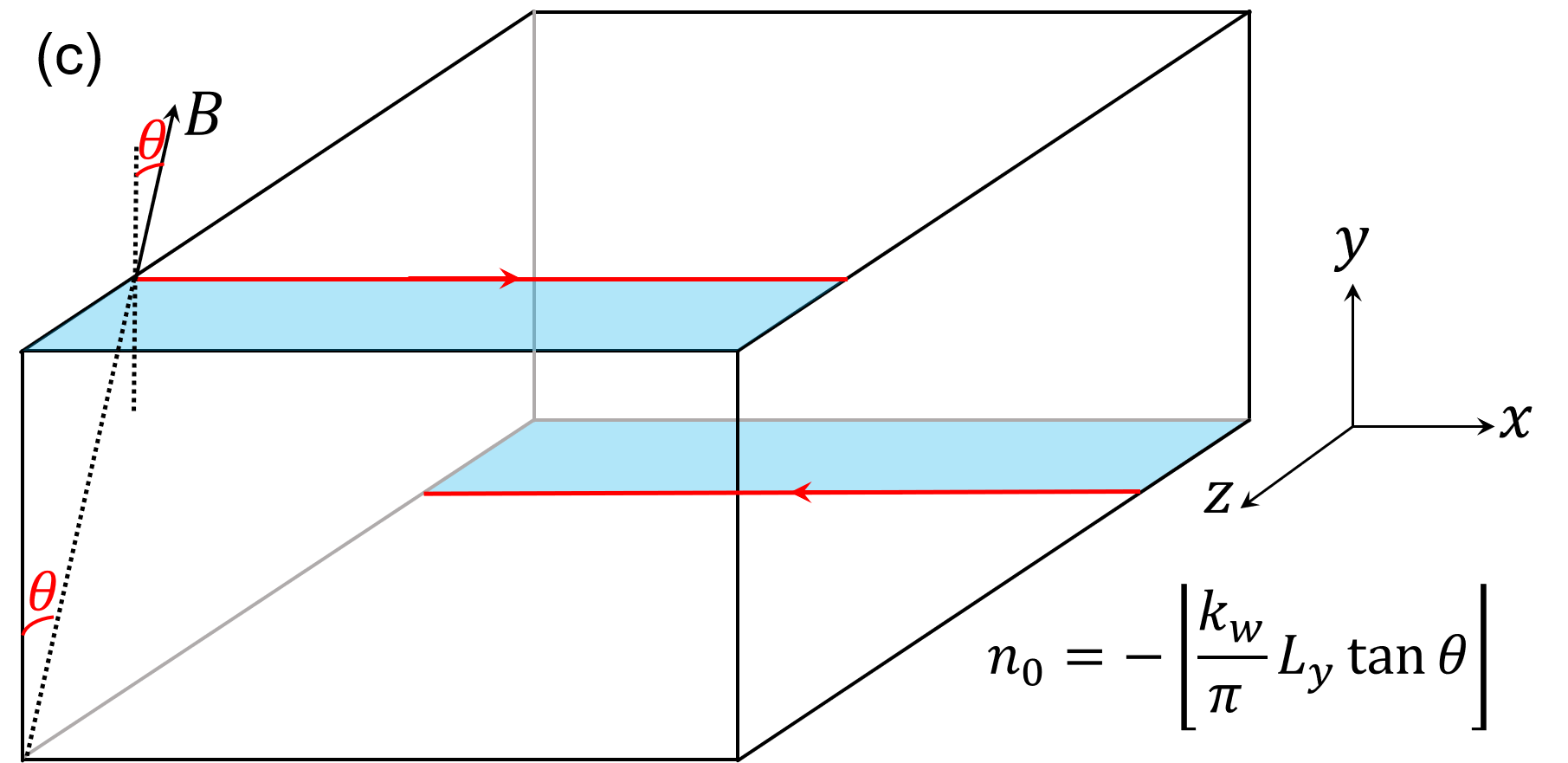}
  \caption{Schematic diagrams of edge states from two different origins. (a) For nonzero $S_k$, edge states are caused by $B_y$ similar to 2D QHE. (b) and (c) show edge states from the intrinsic topological nature of Weyl semimetal under opposite tilting angles $\theta$ and zero $S_k$. }
  \label{hall_formula}
\end{figure}
In the main text, we propose that $\sigma_{xz}^H$ can be expressed as
\begin{equation}
  \sigma_{xz}^H=\frac{e^2}{h}(n_0+n),~~~n=0,-1,-2,...
\end{equation}
It implies two origins where edge states come from. One is similar to 2D QHE denoted by $n$, and the other is similar to quantum anomalous Hall effect (QAHE) denoted by $n_0$. When $\theta=0$, $n_0=0$, i.e. $\sigma_{xz}^H\propto n$. The edge states are distributed like Fig.S~\ref{hall_formula}(a) and Hall conductance is~\cite{SluhaihzouPRL}
\begin{equation}
  \sigma_{xz}^H=-\frac{e^2}{h}\left\lfloor \frac{\hbar S_k}{2\pi e B_y}+\frac{1}{2} \right\rfloor.
\end{equation}
Here, $\left\lfloor... \right\rfloor$ stands for rounding down. Like 2D QHE, $n$ is dependent on $S_k$ and $B_y$.
As in Sec.~S3, when we set $S_k=0$, $n=0$ and $\sigma_{xz}^H\propto n_0$. Hall conductance is expressed as Eq.~\ref{n0} and only depends on the tilting angle $\theta$.

For real materials, $S_k$ is always nonzero. Thus, the observed Hall conductance must be a combination of the above two mechanism. Explicitly, for $\theta>0$, $n\leqslant 0$ and $n_0\leqslant0$. Edge states can be regarded as a superposition of Fig.S~\ref{hall_formula}(a) and Fig.S~\ref{hall_formula}(b), and the coupling of two mechanism will enhance the Hall conductance. For $\theta<0$, $n\leqslant 0$ and $n_0\geqslant0$. The superposition of Fig.S~\ref{hall_formula}(a) and Fig.S~\ref{hall_formula}(c) will make part of edge states localized due to bulk chiral Landau levels like Fig.S~\ref{critical_angle}(b). In this sense, the Hall conductivity satifies Eq.~3 in the main text. Moreover, as in Sec.~S2, it also implies a critical angle $\theta_c$. Once the tilting angle $\theta$ exceeds $\theta_c$, an abvious spatial shift of edge states will occur and subsequently change the sign of $\sigma_{xz}^H$. From the algebraic relation, $n(B_y)\approx n_0(\theta_c)$. Specifically, 
\begin{equation}
  \theta_c\approx -\arctan(\frac{\hbar S_k}{2eB_yk_wL_y}).
\end{equation}
\section{S6. Method to Discretize Hamiltonian}
We start from a continous model, i.e. Eq.~\ref{eqS1}, which decribes a 3D Weyl semimetal with two Weyl nodes.
\begin{equation}
  \begin{aligned} 
    H(\bm{k})=& D_{1} k_{y}^{2}+D_{2}\left(k_{x}^{2}+k_{z}^{2}\right)+A\left(k_{x} \sigma_{x}+k_{y} \sigma_{y}\right) \\ &+M\left(k_{w}^{2}-\bm{k}^{2}\right) \sigma_{z} \end{aligned}
    \label{eqS1}
  \end{equation}
In order to perform numerical calculations, it is necessary to map the continous model into a lattice model by making the following replacements~\cite{SshenshunqingBOOK},
\begin{equation}
  \begin{aligned}
  &k_{i} \rightarrow \frac{1}{a} \sin \left(k_{i} a\right)\\
  &k_{i}^{2} \rightarrow \frac{2}{a^{2}}\left(1-\cos \left(k_{i} a\right)\right)
  \end{aligned}
  \end{equation}
  where $i=x,~y~or~z$ and a is the lattice constant. Then we can do the Fourier transformation to obtain the effective Hamiltonian in the lattice space.
  \begin{equation}
    \begin{aligned}
    a_{\bm{k}}^\dagger=\frac{1}{\sqrt{V}}\sum_{\bm{R}}e^{i\bm{k}\cdot\bm{R}}a_{\bm{R}}^\dagger
    \end{aligned}
    \end{equation}
  Here, $\bm{R}$ denotes the coordinates of the lattice sites. We consequently obtain the final Hamiltonian,
  \begin{equation}
    \begin{aligned}
    H=\sum_{x, y, z} &\left[\left(4 D_{2}+2 D_{1}\right) \sigma_{0}+M\left(k_{w}^{2}-6\right) \sigma_{z}\right] a_{x, y, z}^{\dagger} a_{x, y, z} \\
    &+\left(M \sigma_{z}-D_{1} \sigma_{0}+\frac{i A}{2} \sigma_{y}\right) a_{x, y+1, z}^{\dagger} a_{x, y, z}+h . c . \\
    &+\left(M \sigma_{z}-D_{2} \sigma_{0}\right) a_{x, y, z+1}^{\dagger} a_{x, y+1, z}+h . c . \\
    &+\left(M \sigma_{z}-D_{2} \sigma_{0}+\frac{i A}{2} \sigma_{x}\right) a_{x+1, y, z}^{\dagger} a_{x, y, z}+\text {h.c.}
    \end{aligned}\label{eqS4}
    \end{equation}
  where $\sigma_0$ is the $2\times 2$ identity matrix and $\sigma_{x,y,z}$ are Pauli matrices. In the main text, we set $k_w=1.5$ for numerical calculations. In this case, although Eq.~\ref{eqS4} describes a Weyl semimetal with two Weyl nodes located at $(0,0,\pm 1.7)$ which does not fit $(0,0,\pm 1.5)$ given by Eq.~\ref{eqS1} so well, they both describe Weyl semimetals and gives the same energy at Weyl nodes, $E_w=D_2k_w^2$. Thus, the slight difference between the low-energy model in Eq.~\ref{eqS1} and the lattice model in Eq.~\ref{eqS4} will not affect the global picture of 3D QHE in Weyl semimetals.

  In the presence of an external magnetic field $\bm{B}=(B_1,B_2,B_3)$, Hamiltonian (Eq.~\ref{eqS4}) is modified according to the Peierls substitution. By Landau gauge, the vector potential $\bm{A}=(B_2z-B_3y,-B_1z,0)$, where $\bm{B}=\nabla\times\bm{A}$. Thus, the Peierls substitution is defined by
  \begin{equation}
    \begin{aligned}
    &a_{x+1, y, z}^{\dagger} a_{x, y, z} \rightarrow a_{x+1, y, z}^{\dagger} a_{x, y, z} e^{-i e\left(B_{2} z-B_{3} y\right) / \hbar}\\
    &a_{x, y+1, z}^{\dagger} a_{x, y, z} \rightarrow a_{x, y+1, z}^{\dagger} a_{x, y, z} e^{i e B_{1 z} / \hbar}\\
    &a_{x, y, z+1}^{\dagger} a_{x, y, z} \rightarrow a_{x, y, z+1}^{\dagger} a_{x, y, z}
    \end{aligned}
    \end{equation}
    Finally, the tight binding Hamiltonian of a finite Weyl semimetal is obtained as Eq.~\ref{eq6}.
    \begin{equation}
      \begin{aligned}
      H=\sum_{x, y, z} &\left[\left(4 D_{2}+2 D_{1}\right) \sigma_{0}+M\left(k_{w}^{2}-6\right) \sigma_{z}\right] a_{x, y, z}^{\dagger} a_{x, y, z} \\
      &+\left(M \sigma_{z}-D_{1} \sigma_{0}+\frac{i A}{2} \sigma_{y}\right) a_{x, y+1, z}^{\dagger} a_{x, y, z} e^{i e B_{1} z / \hbar}+h . c . \\
      &+\left(M \sigma_{z}-D_{2} \sigma_{0}\right) a_{x, y, z+1}^{\dagger} a_{x, y+1, z}+h . c . \\
      &+\left(M \sigma_{z}-D_{2} \sigma_{0}+\frac{i A}{2} \sigma_{x}\right) a_{x+1, y, z}^{\dagger} a_{x, y, z} e^{-i e\left(B_{2} z-B_{3} y\right) / \hbar}+h . c .
      \end{aligned}\label{eq6}
      \end{equation}

\section{S7. Local density of states}
We consider an infinitely long Weyl semimetal along $x$ direction under the magnetic field, of which the Hamiltonian is defined by
\begin{equation}
	\begin{aligned}
		H=\sum_{k_x,z,y}&\Bigg[ \left(2D_2\left(2-\cos(k_x+\frac{e}{\hbar}(B_2z-B_3y))\right)+2D_1\right)\sigma_0+A\sin(k_x+\frac{e}{\hbar}(B_2z-B_3y)) \sigma_{x} \\
		  &+M\left(k_{w}^{2}-2\left(3-\cos(k_x+\frac{e}{\hbar}(B_2z-B_3y))\right)\right) \sigma_{z} \Bigg]a_{k_x,z,y}^{\dagger} a_{k_x,z,y}\\
		  & +\Bigg[\left(M\sigma_z-D_1\sigma_0+\frac{iA}{2}\sigma_y\right)a_{k_x,z,y+1}^{\dagger} a_{k_x,z,y}e^{ieB_1z/\hbar} \\
		  & +\left(M\sigma_z-D_1\sigma_0-\frac{iA}{2}\sigma_y\right)a_{k_x,z,y-1}^{\dagger} a_{k_x,z,y}e^{-ieB_1z/\hbar}\Bigg]\\
		  &+\left(M\sigma_z-D_2\sigma_0\right)a_{k_x,z+1,y}^{\dagger} a_{k_x,z,y}+\left(M\sigma_z-D_2\sigma_0\right)a_{k_x,z-1,y}^{\dagger} a_{k_x,z,y}
	 \end{aligned}
 \end{equation}
 In the main text, the external magnetic field is oriented in the $y$-$z$ plane, i.e. $B_1=0$.
 Due to the translational symmetry along $x$ direction, the local density of states of any $y$-$z$ cross section can be easily calculated by recursive Green's function method. By using Dyson equation, the recursive relations between the Green's functions for the $z$th and $z+1$st slice.
 \begin{equation}
   \begin{aligned}
     G_{z+1,z+1}^{(z+1)}&=\left[\bm{Z}-H_{z+1}^0-H_{z,z+1}^\dagger G_{z,z}^{(z)}H_{z,z+1}\right]^{-1}\\
     G_{i,j}^{(z+1)}&=G_{i,j}^{(z)}+G_{i,z}^{(z)}H_{z,z+1}G_{z+1,z+1}^{(z+1)}H_{z,z+1}^\dagger G_{z,j}^{(z)}~~~(i,j\leqq z)\\
     G_{i,z+1}^{(z+1)}&=G_{i,z}^{(z)}H_{z,z+1}G_{z+1,z+1}^{(z+1)}~~~(i\leqq z)\\
     G_{z+1,j}^{(z+1)}&=G_{z+1,z+1}^{(z+1)}H_{z,z+1}^\dagger G_{z,j}^{(z)}~~~(j\leqq z)
   \end{aligned}
 \end{equation}
 Here, $\bm{Z}=(E+i\eta)\bm{I}$ and $\eta$ is an infinitesimal quantity. We choose $\eta=10^{-3}$. $H_{z+1}^0$ is the Hamiltonian of the $(z+1)$st slice and $H_{z,z+1}$ is the hopping matrix between the $z$th and the $z+1$st slice. 
  Consequently, the local density of states at Fermi energy $E_F$ is defined as
 \begin{equation}
   \rho(y,z,E_F)=-\frac{1}{\pi N_y N_z}\sum_{k_x}\operatorname{Im}~tr\left[G(k_x,y,z,E_F)\right]
 \end{equation}
\section{S8. Local current density}
In order to simulate the local current density, a small voltage bias, $V_L-V_R$, is applied between the left and right terminal. Under nonequilibrium Green's function, the local current can be calculated via the time derivative of the electron number operator $\hat{N}_{\bm{i}}$,
\begin{equation}
  \begin{aligned}
  J_{\mathbf{i}} &=e\left\langle\dot{N}_{\mathbf{i}}\right\rangle \\
  &=\frac{i e}{\hbar}\left\langle\left[H, \sum_{\alpha} N_{\mathbf{i} \alpha}\right]\right\rangle \\
  \end{aligned}
  \end{equation}
After some simple derivations, the local currents between the neighbouring sites $\bm{i}$ and $\bm{j}$ can be calculated as 
\begin{equation}
  \begin{aligned}
  J_{\mathrm{i} \rightarrow \mathrm{j}}=& \frac{2 e}{h} \sum_{\alpha, \beta} \int_{-\infty}^{e V_{R}} d E \operatorname{Im}\left\{H_{\mathrm{i} \alpha, \mathrm{j} \beta}\left[G^{r}\left(\Gamma_{L}+\Gamma_{R}\right) G^{a}\right]_{\mathrm{j} \beta, \mathrm{i} \alpha}\right\} \\
  &+\frac{2 e^{2}}{h} \sum_{\alpha, \beta} \operatorname{Im}\left[H_{\mathrm{i} \alpha, \mathrm{j} \beta} G_{\mathrm{j} \beta, \mathrm{i} \alpha}^{n}\left(E_{F}\right)\right]\left(V_{L}-V_{R}\right)
  \end{aligned}
  \end{equation}
  where $V_{L},~V_{R}$ are the voltages at the leads. $G^r(G^a)$ is the retarded (advanced) Green's function and we choose $\eta=10^{-9}$. $H_{\bm{i},\bm{j}}$ is the coupling Hamiltonian between sites $\bm{i}$ and $\bm{j}$. $\Gamma_L$($\Gamma_R$) is the linewidth function of left(right) lead. Moreover, $\Gamma_{L/R}=i\left[\Sigma_{L/R}^r-\Sigma_{L/R}^a\right]$ where $\Sigma_{L/R}^r$ is the self-energy and $\Sigma_{L/R}^a={\Sigma_{L/R}^r}^\dagger$.
\section{S9. Hall conductance of a Weyl semimetal slab}
\begin{figure}[htbp]
  \centering
  \includegraphics[width=.5\columnwidth]{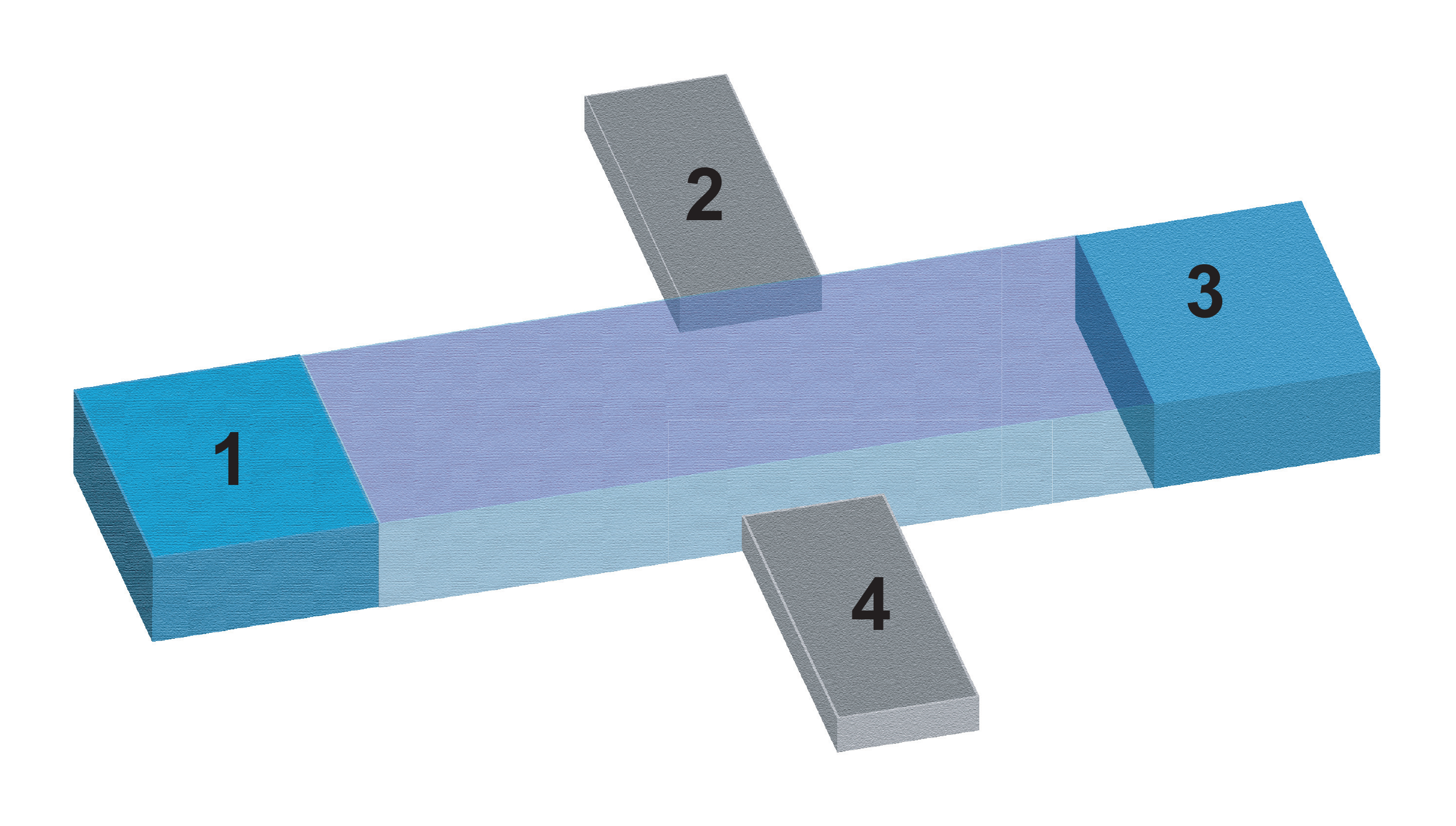}
  \caption{A schematic diagram of the four-terminal device in main text. Lead 1 and 3 are Weyl semimetals, while lead 2 and 4 are normal metals.  }\label{four-terminal}
  \end{figure}
  The four-terminal device we use in the main text is depicted in Fig.S~\ref{four-terminal}. According to $Landaur-B\ddot{u}ttiker$ formula, the voltage and current probes in this four-terminal device satisfy:
  \begin{equation}
    I_i=\frac{e^2}{h}\sum_{j\neq i}T_{ij}\left[V_i-V_j\right]\label{eq12}
  \end{equation} 
  where $T_{ij}$ is the transmission coefficient betweem inter-terminal $i$ and $j$. $V_i$ is the voltage of the $i$th terminal. According nonequilibrium Green's function methos, the transmission coefficient is calculated By
  \begin{equation}
    T_{ij}=\operatorname{tr}\left[\Gamma_iG^r\Gamma_jG^a\right].
  \end{equation}
  Here, we choose $\eta=10^{-9}$. We can also write Eq.~\ref{eq12} into a matrix form
  \begin{eqnarray}\label{eq13}
    \begin{bmatrix}
      I_1\\
      I_2\\
      I_3\\
      I_4
    \end{bmatrix}=\frac{e^2}{h}
                        \begin{bmatrix}
                          T_{12}+T_{13}+T_{14} & -T_{12} & -T_{13} & -T_{14} \\
                          -T_{21} & T_{21}+T_{23}+T_{24} & -T_{23} & -T_{24} \\
                          -T_{31} & -T_{32} & T_{31}+T_{32}+T_{34} & -T_{34} \\
                          -T_{41} & -T_{42} & -T_{43} & T_{41}+T_{42}+T_{43} \\
                        \end{bmatrix}
                        \begin{bmatrix}
                          V_1\\
                          V_2\\
                          V_3\\
                          V_4
                        \end{bmatrix},
  \end{eqnarray} 
  Without loss of generality, we set $V_3$ zero. Combined with Kirchhoff's current law, i.e., $I_1+I_2+I_3+I_4=0$, Eq.~\ref{eq13} can be reuced into:
  \begin{eqnarray}\label{eq15}
    \begin{bmatrix}
      I_1\\
      I_2\\
      I_4
    \end{bmatrix}=\frac{e^2}{h}
                        \begin{bmatrix}
                          T_{12}+T_{13}+T_{14} & -T_{12} & -T_{14} \\
                          -T_{21} & T_{21}+T_{23}+T_{24} &  -T_{24} \\
                          -T_{41} & -T_{42} &  T_{41}+T_{42}+T_{43} \\
                        \end{bmatrix}
                        \begin{bmatrix}
                          V_1\\
                          V_2\\
                          V_4
                        \end{bmatrix}.
  \end{eqnarray} 
  For a Hall measurement, $I_2=I_4=0$. Here, we define a $3\times 3$ matrix $A$ as
  \begin{eqnarray}
                        \begin{bmatrix}
                          T_{12}+T_{13}+T_{14} & -T_{12} & -T_{14} \\
                          -T_{21} & T_{21}+T_{23}+T_{24} &  -T_{24} \\
                          -T_{41} & -T_{42} &  T_{41}+T_{42}+T_{43} \\
                        \end{bmatrix}^{-1}.
  \end{eqnarray}
Thus the Hall conductance is calculated through:
\begin{equation}
  \sigma_{xz}^H=\frac{e^2}{h}\frac{1}{A_{21}-A_{31}}.
\end{equation}

\begin{thebibliography}{41}
\bibitem [{ber()}]{berrycurvature_expla}%
    \BibitemOpen
    \href@noop {} {\enquote {\bibinfo {title} {Due to the quantization of chiral
    bulk landau levels, bulk electrons will not feel the berry curvature. on
    nontrivial surfaces, berry curvature on fermi arcs is so small that it can be
    neglected. in particular, berry curvature of the fermi arcs in our model is
    zero~\cite{luhaihzouPRL}.}}\ }\BibitemShut {NoStop}%
\end{thebibliography}
%
  
    \end{widetext}
\end{document}